\documentclass[twocolumn]{aastex61}
\usepackage{natbib}
\usepackage{amsmath}
\usepackage{graphicx}
\submitjournal{\emph{The Astrophysical Journal}}
\received{2018, February 17}
\revised{2018, October 8}
\accepted{2018, October 17}
\published{2018, November 30}

\shorttitle{Megaparsec-scale Spiral in A1763 }
\shortauthors{Douglass et al.}

\begin{document}

\title{The Megaparsec-scale Gas-sloshing Spiral in the Remnant Cool Core \\ Cluster Abell 1763 }

%\correspondingauthor{E. M. Douglass}
%\email{edmund.douglass@farmingdale.edu}
\correspondingauthor{E. M. Douglass}
\email{douglae@farmingdale.edu}

\author{E. M. Douglass}
\affil{Farmingdale State College - SUNY, 2350 Broadhollow Rd., Farmingdale, NY 11735}
\affil{American Museum of Natural History,  Central Park West and W. 79 St, New York, NY 10024}

\author{E. L. Blanton}
\affil{Boston University, 725 Commonwealth Ave, Boston, MA 02215}

\author{S. W. Randall }
\affiliation{Harvard Smithsonian - Center for Astrophysics, 60 Garden St. Cambridge, MA 02138}

\author{T. E. Clarke}
\affiliation{Naval Research Laboratory, 4555 Overlook Ave SW, Washington, DC 20375}

\author{L. O. V. Edwards}
\affiliation{California Polytechnic State University - San Luis Obispo, CA 93407}

\author{Z. Sabry}
\affil{Farmingdale State College - SUNY, 2350 Broadhollow Rd., Farmingdale, NY 11735} 
\affil{Swarthmore College, 500 College Ave, Swarthmore, PA 19081} 

\author{J. A. ZuHone}
\affiliation{Harvard Smithsonian - Center for Astrophysics, 60 Garden St. Cambridge, MA 02138}

\begin{abstract}

We present a multiwavelength study of the massive galaxy cluster Abell 1763 at redshift z = 0.231.  Image analysis of a 19.6 ks \emph{Chandra} archival observation reveals a cluster-wide spiral of enhanced surface brightness in the intracluster medium (ICM). While such spirals are understood to form in clusters with sloshing strong cool cores (SCCs), the gas comprising the spiral's apex is of intermediate entropy ($\sim$ 110 keV cm$^{2}$) and cooling time ($\sim$ 6.8 Gyr), indicating core disruption is occurring throughout the spiral formation process. Two subclusters dominated by the second- and third-ranked galaxies in the system lie along a line parallel to the elongation axis of the primary cluster's ICM. Both subsystems appear to have fallen in along a previously discovered intercluster filament and are each considered candidates as the perturber responsible for initiating disruptive core sloshing. Dynamical analysis indicates infall is occurring with a relative radial velocity of $\sim$ 1800 km s$^{-1}$. The brightest cluster galaxy of Abell 1763 possesses a high line-of-sight peculiar velocity (v$_{pec}$ $\sim$ 650 km s$^{-1}$) and hosts a powerful (P$_{1.4}$ $\sim$ 10$^{26}$ W Hz$^{-1}$) bent double-lobed radio source, likely shaped by the relative bulk ICM flow induced in the merger. The cluster merger model of SCC destruction invokes low impact parameter infall as the condition required for core transformation. In contrast to this, the high angular momentum event occurring in Abell 1763 suggests that off-axis mergers play a greater role in establishing the non-cool core cluster population than previously assumed.      
\end{abstract}
\section{Introduction} \label{sec:intro}

With diameters spanning several megaparsecs, galaxy clusters provide us with the unique opportunity to observe the forces that shape our universe acting across the grandest scales.  Formed through the hierarchical assembly of smaller galaxy groups, clusters grow to become the largest gravitationally bound systems in nature. The majority of their baryonic mass is found in the hot diffuse plasma (intracluster medium, ICM) through which their member galaxies orbit \citep[see][]{Sarazin88}.  Outweighing the galactic stellar mass by a factor of 5 to 1, the ICM is heated to temperatures of tens of millions of degrees by the large gravitational potential of the systems (with masses 10$^{14}$ M$_{\odot}$ $<$ M $<$ 10$^{15}$ M$_{\odot}$).  The ICM preserves clues to clusters' evolutionary histories as large-scale mergers, lesser accretion events, AGN outbursts, and relaxation all leave observable imprints on the spatial distribution and thermal characteristics of the gas.   

As the ICM shines via thermal bremsstrahlung ( L$_X \propto$ n$^2$ T$^{1/2}$), systems with the densest cores can radiate away energy at a rate high enough to drive down their central temperatures to a fraction of the average global value \citep[see][]{Fabian94}. This results in the formation of a ``cool core" (CC) of low-entropy gas. Traditionally, galaxy clusters are divided into two groups: those that have CCs and those that do not (non-cool cores, NCCs). A multitude of X-ray derived diagnostics are used to distinguish between CC and NCC clusters, including central cooling time, core entropy, central temperature gradient, surface brightness cuspiness, and mass deposition rate. Different diagnostic systems may classify systems of intermediate core characteristics on different sides of the CC/NCC divide. Consequently, the line demarcating the CC and NCC cluster populations is a blurry one.

In a detailed analysis of the X-ray properties of the cores of clusters comprising the \emph{HIFLUGCS} cluster sample, \citet{Hudson10} identify central cooling time (t$_{cool}$) as the best parameter to differentiate between CCs and NCCs. The cooling time threshold of t$_{cool}$ $<$ 7.7 Gyr is chosen as an appropriate break value as it corresponds to the look-back time for z = 1, the time span over which most clusters could relax and form a CC. When central entropy is considered as a CC diagnostic, the CC and NCC cluster populations segregate into the same t$_{cool}$ determined subpopulations at a break value of K$_{0}$ = 150 keV cm$^2$.

The CC class is further subdivided into strong-CCs (SCCs) and weak-CCs (WCCs). SCCs have short central cooling times (t$_{cool}$ $<$ 1 Gyr) and low central entropies (K$_{0}$ $\lesssim$ 30 keV cm$^2$) while WCCs have core properties intermediate to the SCC and NCC classes (1 Gyr $<$ t$_{cool} <$ 7.7 Gyr, 30 keV cm$^2 <  K_{0} <$ 150 keV cm$^2$). Within their statistically complete flux-limited sample of 64 clusters, 44\% have SCCs while the remaining 56\% is split evenly between WCCs and NCCs.

The origin of the CC/NCC population distribution is not well understood but is assumed to result from differing evolutionary histories of the clusters.  It is expected that an SCC should form in collapsed massive gas-rich halos in the absence of additional energy injection into the ICM \citep[see][]{Sarazin88}. From this, it follows that something must have happened to NCC clusters to make them what they are. Cluster mergers and AGN outbursts are considered the most likely mechanisms by which an NCC is produced, but the relative contribution by each to populating the NCC class is uncertain. Furthermore, regardless of the mechanism of transformation, it is uncertain whether the transition is permanent \citep{Rossetti11,RM10}.     

The merger model for dynamically driven NCC formation is supported by observational evidence that unlike NCCs, SCC clusters generally have smooth, axisymmetric surface brightness distributions \citep{OHara06, Maughan12}. In agreement with this, numerous WCCs are found in clusters displaying signatures of ongoing major infall events. The cores of these systems are often referred to as ``remnant cool cores," as they had most likely previously been in an SCC state. These systems appear to be experiencing head-on mergers, hosting disrupted CCs displaced from their brightest cluster galaxies (BCGs), e.g. Abell 1758 \citep{David04}, Abell 2146 \citep{Russell10}, and Abell 3562 \citep{Finoguenov04}.  Recent cosmological simulations \citep[e.g.,][]{Hahn17} suggest high energy, low angular momentum mergers, such as those transpiring in the above systems, are required for the disruption of SCCs and the formation of NCC clusters.

While high-energy head-on collisions may disrupt SCCs, simulated off-axis mergers are often found to merely perturb them \citep[][for an exception, see \citet{Zuhone11}]{Tittley05,AM06}.  The off-axis passage of a subcluster can lead to a slight displacement between the primary system's core gas and its dark matter (DM) peak. This sets up oscillatory motion of the core in a process termed `sloshing' \citep{Markevitch01}.  Core sloshing is the accepted mechanism responsible for the production of cool, low-entropy gas spirals \citep{MV07} observed in the centers of many seemingly relaxed systems (e.g. \citet{Clarke04}, \citet{Rossetti07}, \citet{Lagana10}, \citet{Blanton11}, \citet{Paterno13}, \citet{Ghizzardi14}).

An alternative to the cluster merger model for NCC formation invokes powerful outbursts by the centrally located AGN to inject the energy necessary to significantly disrupt ICM cooling. An early iteration of this model was driven by cosmological simulations of \citet{McCarthy05}, which failed to produce NCCs in the absence of non-gravitational heating. The authors concluded that the formation of NCCs requires a non-gravitational preheating \citep{Kaiser91,EH91} of embryonic CCs to destroy the seeds from which SCCs would normally grow.  The simulations of \citet{Poole08} reinforced this conclusion as their SCCs, once established, were extremely difficult to destroy via individual dynamical events.  \citet{Burns08} showed that significant mergers occurring during the earliest epochs of cluster formation could produce long-lived NCCs. However, they also concluded, like the aforementioned studies, that established SCCs were nearly invulnerable to dynamically driven destruction.  

The preheating scenario has been directly challenged recently by the findings of \citet{Iqbal17}, which in a study of the outskirts of a large sample of clusters, ruled out evidence for non-gravitational preheating to high significance. Focusing on more evolved clusters, the simulations of \citet{GM10} show that AGN outbursts of extreme energies (10$^{62}$ erg) similar to those seen in RX J1532.9+3021 \citep{Hlavacek13} and MACS J1931.8-2634 \citep{Ehlert11} would be capable of destroying more established SCCs in the low-z universe. However, the few clusters that host such powerful outbursts are the site of some of the strongest SCCs observed to date ($\dot{M}$ $>$ 500 M$_{\odot}$ yr$^{-1}$). Though their cores are disrupted by the outbursts, they do not appear to be transitioning to an NCC state. 

In contrast to the indestructible cores of the simulations described above, SCCs produced in more recent cosmological simulations \citep{Rasia15, Hahn17} can be destroyed via high-energy, low angular momentum mergers. These results can in a large part be attributed to more accurate treatments of the hydrodynamics in the simulations. The smooth particle mesh (SPH) scheme used by \citet{Rasia15}, unlike the SPH scheme of \citet{Poole08}, appears to handle turbulence and fluid instabilities properly, allowing gas mixing and the formation of higher entropy cores \citep[see][]{Agertz07,Mitchell09,Zuhone11}, while the higher resolution (3.8 h$^{-1}$ kpc) of the Eulerian code of \citet{Hahn17} allows gas mixing in a way prevented by the more coarse resolution ($\sim$16 h$^{-1}$ kpc) of the \citet{Burns08} code. Of no small significance, the simulations of both \citet{Rasia15} and \citet{Hahn17} include effects of AGN feedback, likely resulting in cores that are more susceptible to dynamically driven destruction.                  

These recent cosmological simulations show that low angular momentum mergers are capable of destroying SCCs at a frequency consistent with the observed CC-NCC distribution, however, observational studies remain inconclusive as to whether mergers or AGN are primarily responsible for the CC-NCC transition \citep[e.g.][]{Medezinski17}, or if both play meaningful roles in populating the NCC class. It is therefore important to examine intermediate objects that straddle the defining though somewhat blurry line \citep{Hudson10,AndradeSantos17}, between CC and NCC cluster classifications.  In studying these systems, we gain insight into the phenomena responsible for a cluster's transition from one class to another and the origin of the CC/NCC distribution itself.        

In this paper we present an in-depth multiwavelength analysis of such a system, the massive (M$_{200}=1.7 \times 10^{15} M_{\odot}$, \citet{Rines13}) galaxy cluster Abell 1763 at redshift z=0.23. It lies at the southwestern end of an intercluster filament which extends toward a smaller cluster (Abell 1770 at z=0.22), 12 Mpc (projected) to the northeast \citep{Edwards10,Fadda08}.  A NE-SW elongation of the X-ray surface brightness distribution of Abell 1763 is consistent with infall along this filament. It hosts a dual peaked ICM core and displays significant radial velocity substructure (including a $\Delta$v $\sim$ 1800 km s$^{-1}$ merger and a high peculiar velocity BCG, v$_{pec}$ $\sim$  - 650 km s$^{-1}$).  As a likely consequence of the merger event causing these observed phenomena, an ICM spiral has formed, winding outward from the core, which is detectable as an excess in the cluster's two-dimensional surface brightness distribution to a radius of 850 kpc.  Despite the presence of the spiral, the core possesses a central cooling time and entropy well above the SCC limit (t$_{cool}$ $\sim$ 6.8 Gyr, K$_0$ $\sim$ 110 keV cm$^2$). As spiral formation indicates high impact parameter infall \citep{AM06}, Abell 1763 is an unexpected instance of CC disruption occurring in an off-axis (rather than head-on) merger environment.

A member of the ROSAT Brightest Cluster Sample Survey of \citet{Ebeling98}, Abell 1763 is classified as a B­M Type III cluster \citep{BM70} of Abell richness class 3 \citep{Abell58}. The BCG (r$_{sdss}$=16.39) is host to a powerful (P$_{1.4GHz}$ = 1.1 $\times 10^{26}$ W Hz$^{-­1}$ , \cite{OL97} bent double-­lobed radio source displaying the typical jet-­hotspot-­lobe morphology of wide angle tail (WAT) radio sources \citep{OR76, ODon90}. The cluster was observed for 19.6 ksec by Chandra during Cycle 4 (PI: Van Speybroeck). The observation has been included in a number of large cluster studies \citep[e.g.][]{Cavagnolo09, Ehlert15, Zhu16} and the cluster is often identified as an NCC system.  However, the analysis outlined in this paper indicates that a more nuanced interpretation of its state may be required, as we appear to be observing what was until very recently an SCC.   All uncertainties are given to 1-sigma confidence intervals. Quoted SDSS magnitudes are de Vaucouleurs  model magnitudes (\emph{devMag}).  We assume  $\Lambda$=0.7, $\Omega_m$=0.3 and H$_0$ = 70 km s$^{-1}$ Mpc$^{-1}$. At z = 0.23, the angular diameter distance is D$_A$ = 757.9 Mpc, the luminosity distance is D$_L$ =  1146.6 Mpc, and 1$\arcsec$=3.67 kpc.

%%%%%%%%%%%%%%%
%%%%%%%%%%%%%%%

\section{OBSERVATIONS AND DATA REDUCTION} \label{sec:obs}
\subsection{X-ray Data}
The \emph{Chandra} observation of Abell 1763 (observation ID: 3591, PI: Van Speybroeck) was carried out using the Advanced CCD Imaging Spectrometer (ACIS) in 2003 August.  The cluster was observed with the ACIS-I array for 19.6 ks in \texttt{VFAINT} mode.  The event data were recalibrated and reprocessed using the CIAO (version 4.7) software package, including the appropriate gain maps and calibration files (CALDB version 4.6.5).  The \texttt{chandra\textunderscore repro} script within CIAO was used to create a new bad pixel file and level=2 event file.   We analyzed light curves on the I1 chip throughout the observation period.  The absence of any 3-$\sigma$ deviations above average count rate during the observation led to the conclusion that no background flares occurred to contaminate the data.  

Background files were created using the appropriate blank-sky files of M. Markevitch included in CALDB and reprojected to match the observation of Abell 1763.  The background fields were normalized to the observation by the ratio of counts between the source and blank-sky files in the 10-12 keV energy range.  An exposure map was created using standard techniques outlined in the CXC CIAO analysis guides.  Point sources were detected by applying the CIAO wavelet detection tool \emph{wavdetect} to an unbinned, unsmoothed, exposure-corrected, background-subtracted image.  Wavelet scales of 1, 2, 4, 8, and 16 were used. The point sources were removed for spectral analysis. For imaging analysis, the \emph{dmfilth} tool was applied to replace point sources with pixel values interpolated from the surrounding medium. 

\subsection{Radio Data}

Radio observations of Abell 1763 were obtained from the NRAO Karl G. Jansky Very Large Array (hereafter VLA) archive program 15A-230 (observed 2015 March 6). Observations were taken in the VLA B configuration with a total time of 28 minutes on target. The data were calibrated within CASA version 4.5.3 following the standard continuum data reduction procedures. The bandpass normalization and flux scale were set using the calibrator 3C286. Initially, the data were Hanning smoothed to reduce the effects of strong radio frequency interference (RFI). We applied the predetermined elevation-dependent gain curves and antenna position corrections. We undertook an initial round of calibration on 3C286 to determine the delay solutions and an initial bandpass solution. We applied these to the data and subsequently used the CASA automated rflag routine to identify and flag RFI. Following the flagging of RFI, we redetermined the delays and bandpass using 3C286 and flagged again using rflag to identify weaker RFI remaining in the data. We set the flux scale of 3C286 using values from \citet{PB13}. Following calibration of the data (including the secondary calibrator J1327+4326), we split off the target field applying the calibration and undertook a final round of RFI flagging using rflag to identify any remaining low-level interference in the target field.

We imaged the target in CASA using w-projection with 175 planes in order to take into account the non-coplanar nature of the array. We used \citet{Briggs95} weighting with robust factor of 0 and accounted for source spectral index during deconvolution using a Taylor term expansion with nterms=2 \citep{RC11}. In addition, we used multifrequency synthesis to take into account effects from the wide bandwidth of the VLA. Following the initial target imaging, we undertook three rounds of iterative phase-only self-calibration to improve the quality. Clean masks were manually set by running the initial clean in an interactive mode, and those clean masks were updated in subsequent imaging. The final image rms is 31 $\mu$Jy/beam with a beam of 3.6$\times$3.0 arcsec at a position angle of -69$^\circ$. 

\subsection{Galaxy Redshifts}
Galaxy velocity data used in the dynamical analysis presented in Section \ref{sec:velocities} is primarily composed of publicly available redshifts collected as part of the Hectospec-Cluster Survey \citep[HeCS,][]{Rines13}.  We include all galaxies that were considered to be cluster members in \citet{Rines13}, falling within their determined value of r$_{200}$ (r$_{200}$ = 2.3 Mpc). r$_{200}$ is the radius inside of which the average density is 200 times the critical density $\rho_c(z)$ at the redshift of the cluster.  These 112 galaxy velocities were supplemented with redshifts of an additional 25 galaxies within r$_{200}$  available in the Sloan Digital Sky Survey Data Release 13 \citep[SDSS - DR13,][]{Albareti17}.  

%%%%%%%%%%%%%%%
%%%%%%%%%%%%%%%

\section{SPATIAL DISTRIBUTION OF CLUSTER EMISSION} \label{sec:spatial}
\begin{figure}
\begin{center}
\hspace*{-2cm}  
\includegraphics[width=200mm]{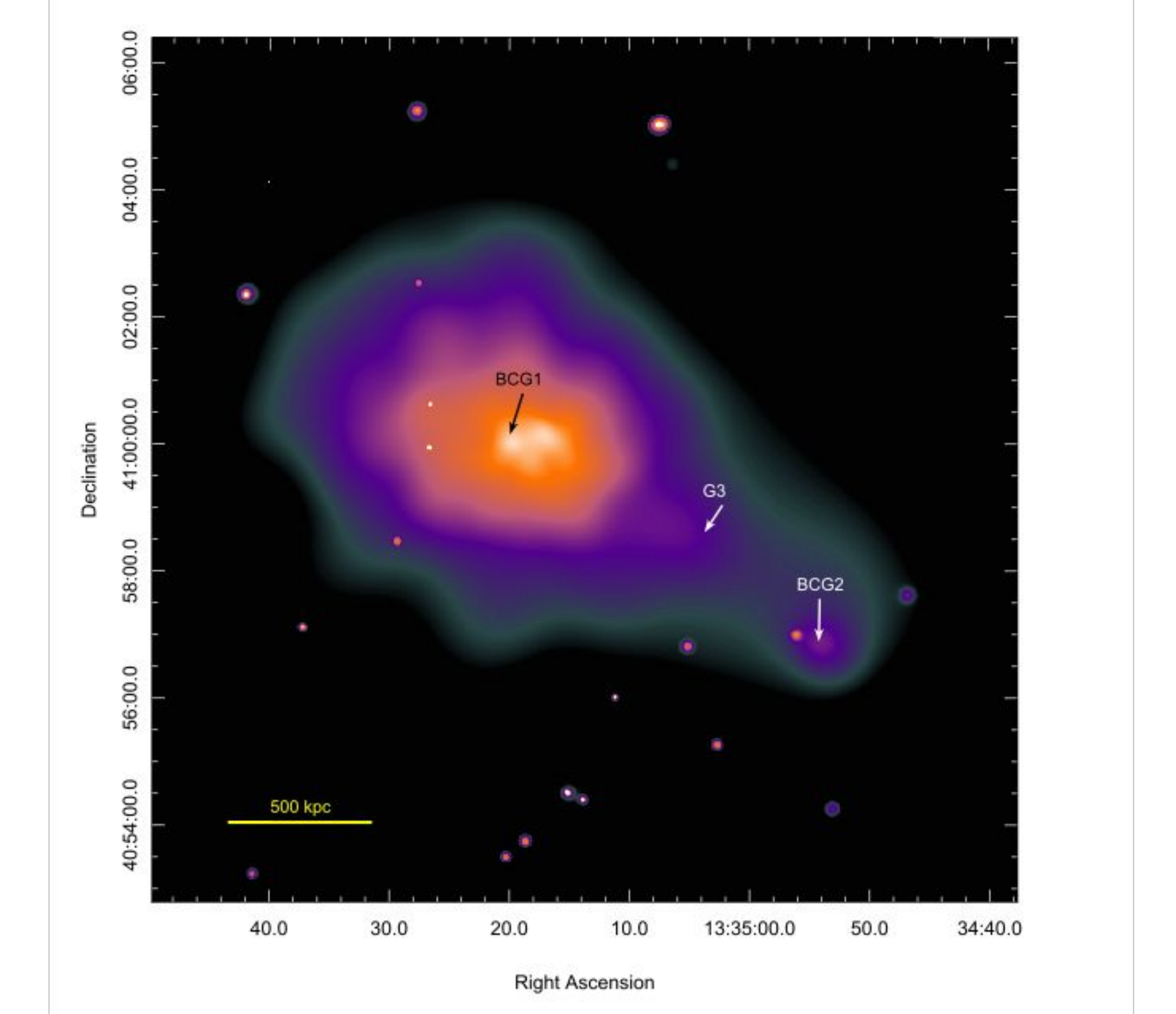}
\caption{Adaptively smoothed Chandra ACIS-I image of Abell 1763 in the energy range of 0.5-7.0 keV. Large-scale distribution is elongated in the NE-SW direction.  A subcluster lies 1.2 Mpc to the southwest of the primary cluster. The three brightest galaxies in the system (BCG1, BCG2, and G3) are labeled. Image is 3 Mpc (13.6$'$) on a side. }
\label{fig:csmooth3Mpc}
\end{center}
\end{figure}

An adaptively smoothed 0.5 - 7.0 keV \emph{Chandra} X-ray image of Abell 1763, created using the \emph{csmooth} task in CIAO (minimum signal-to-noise ratio of 4, maximum of 5), is shown in Figure \ref{fig:csmooth3Mpc}.  The smoothed emission reveals a primary distribution of gas elongated in the NE-SW direction accompanied by a smaller concentration of gas, or subcluster, located a projected distance of 1.2 Mpc to the southwest.  The primary cluster and subcluster are connected by a lower luminosity bridge of emission.

\begin{figure}
\begin{center}
\hspace*{-1cm}  
\includegraphics[width=200mm]{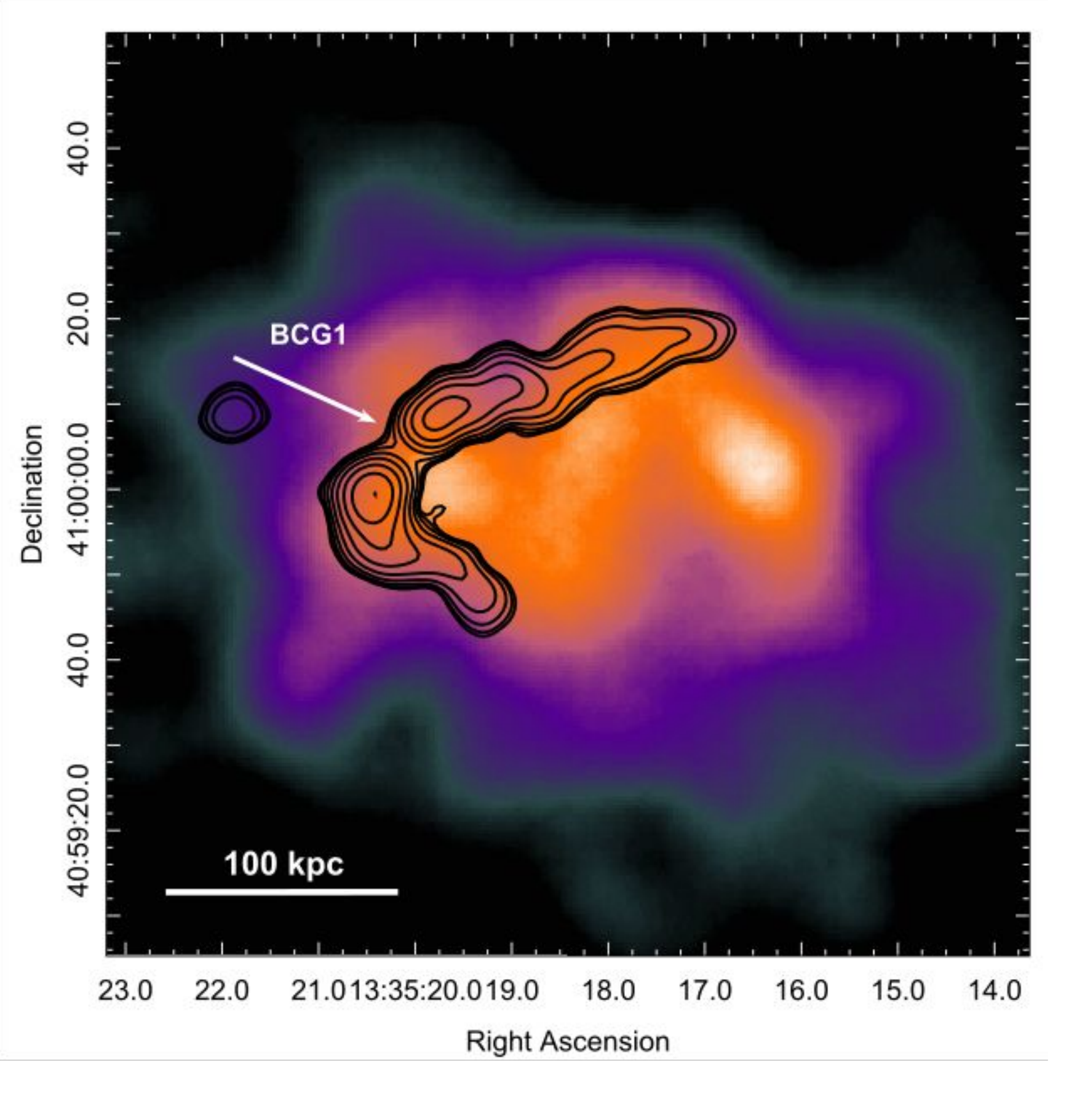}
\caption{Gaussian-smoothed \textbf{($\sigma$ = 5$\arcsec$)} image of the primary cluster core in the energy range of  0.5 - 7.0 keV.  VLA 1.4 GHz contours of WAT source 1333+412 are overlaid in black.  The image is 400 kpc on a side.   Two emission peaks of near equal brightness can be seen within the core region.  The line bisecting the opening angle of the WAT is nearly parallel to the ICM elongation axis. The rms level for the 1.4 GHz image is 31 $\mu$Jy/beam with a beam of 3.6$\times$3.0 arcsec at a position angle of -69$^\circ$. The contours increase logarithmically from 30 times the rms level.}
\label{fig:csrad}
\end{center}
\end{figure}

\begin{figure}
\begin{center}
\hspace*{-1cm}  
\includegraphics[width=200mm]{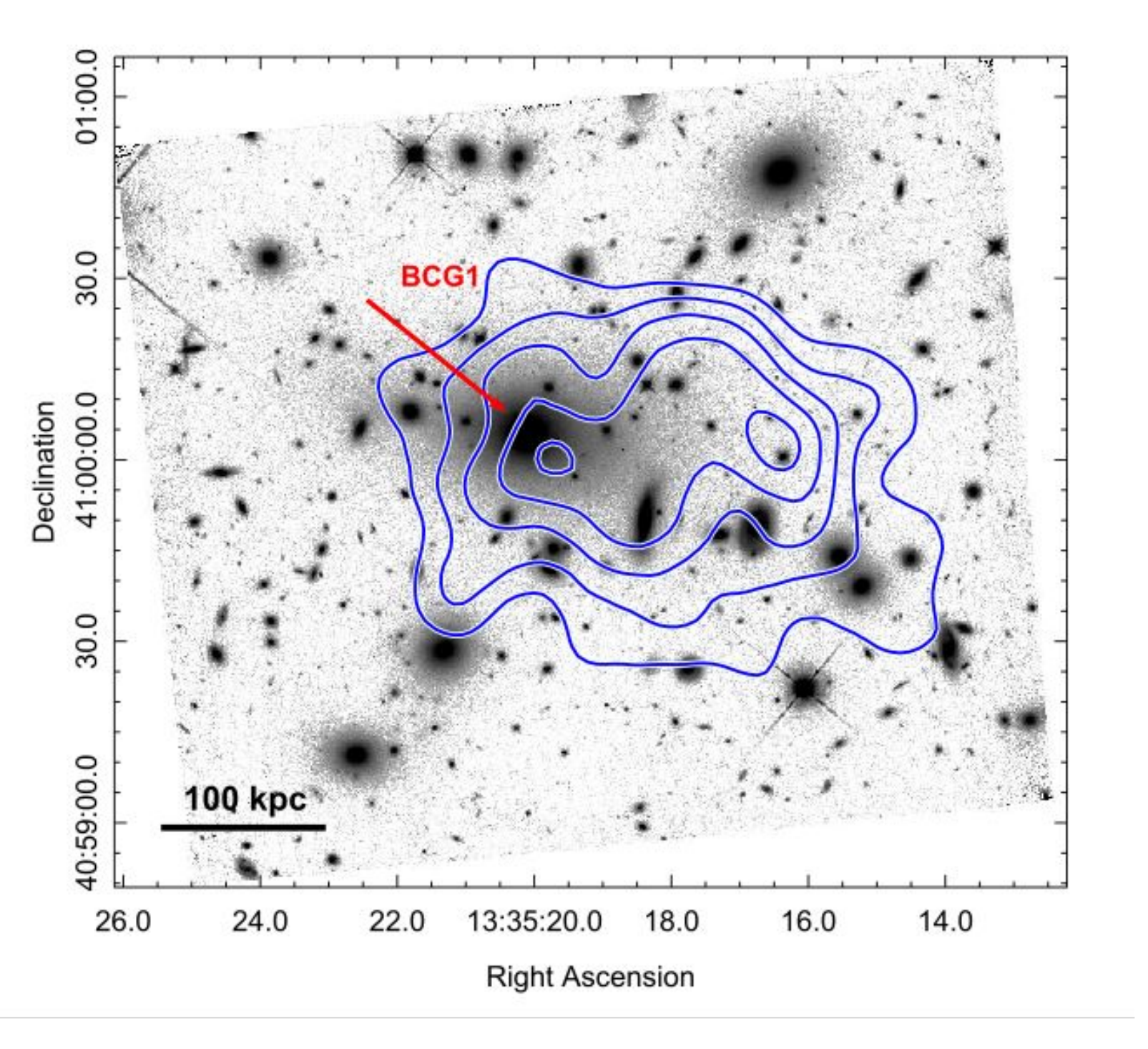}
\caption{\emph{Hubble Space Telescope} image of the central 400 kpc of primary cluster. Contours of X-ray emission from Fig. \ref{fig:csrad} are overlaid.  The eastern emission peak is nearly coincident with BCG1, while the western emission peak has no optical counterpart \textbf{of comparable brightness to BCG1}.  }
\label{fig:csHST}
\end{center}
\end{figure}

Close examination of the central 400 kpc of the core (Figure \ref{fig:csrad}) reveals two emission peaks of approximately equal brightness separated by $\sim$ 140 kpc.  The eastern peak is nearly coincident with the BCG (with SDSS \emph{r-mag} m$_r$=16.40). Its center lies 15 kpc to the southwest of the galaxy's center (Figure \ref{fig:csHST}). The western peak is located 150 kpc due west of the BCG lacking its own BCG-type counterpart. 

The line bisecting the opening angle of the WAT hosted by the BCG \citep[shown here in Figure \ref{fig:csrad} and in][]{Edwards10} is roughly parallel to the major axis of ICM elongation. This is consistent with the NE-SW direction being the primary plane-of-sky axis along which a dynamical event is occurring \citep{Gomez97}.  Similar to the cool wake extending from the galaxy host of 4C 34.16 \citep{Sakelliou05} and the high metallicity excess trailing behind the BCG host of 0647+693 in Abell 562 \citep{Douglass11}, the presence of an overdensity of emission between the lobes of bent WAT sources, aligned to the global ICM elongation, provides insight into the ICM velocity field in the vicinity of the BCG.  

The secondary concentration of emission lies 1.2 Mpc to the southwest of the primary cluster along the elongation axis (Figure \ref{fig:csmooth3Mpc}).  Situated at its center is the second-ranked galaxy in the Abell 1763 system (r$_{sdss}$=16.63), also host to a WAT (see Section \ref{sec:disc}).  Henceforth, we will refer to the brightest galaxies in the primary cluster and subcluster as BCG1 and BCG2, respectively.  

The surface brightness distribution of the primary system (see Figure \ref{fig:csmooth3Mpc}) includes a tongue of emission extending toward the BCG2 subcluster along the elongation axis of the large-scale ICM.  Located within this emission feature, roughly halfway between BCG1 and BCG2, lies the third-ranked galaxy in the Abell 1763 complex (hereafter referred to as G3). A recent study presented in \citet{Haines18} probes \emph{XMM} and Subaru observations of the LoCUSS galaxy cluster sample in search of infalling subsystems. They identify both the BCG2 and G3 systems as subclusters infalling on Abell 1763. They calculate X-ray luminosity derived subcluster masses on the order of 10$^{14}$ M$_{\odot}$ for each.

\subsection{Cluster Substructure}
\subsubsection{Model-subtracted Residual Emission}
A 2-D $\beta$-model  (Sherpa model: \texttt{beta2d}) was fitted to a Gaussian-smoothed image of the cluster ($\sigma = 1"$) with center, core radius, ellipticity, position angle, amplitude, and power-law index free to vary.  The BCG2 subcluster was excluded from the data set for fitting.  The elliptical fit returns a core radius of r$_c$=183 $_{-40}^{+52}$ kpc, an ellipticity of e=0.30 $_{-0.10}^{+0.09}$, a major axis position angle of $\theta$ = 74$^o$ $\pm$ 11$^o$ (measured east of north), and power-law index of $\alpha$ = 1.14 $_{-0.16}^{+0.25}$. The center of the fit falls slightly closer to BCG1 between the two central X-ray emission peaks.

\begin{figure}
\begin{center}
\hspace*{-1cm}  
\includegraphics[width=200mm]{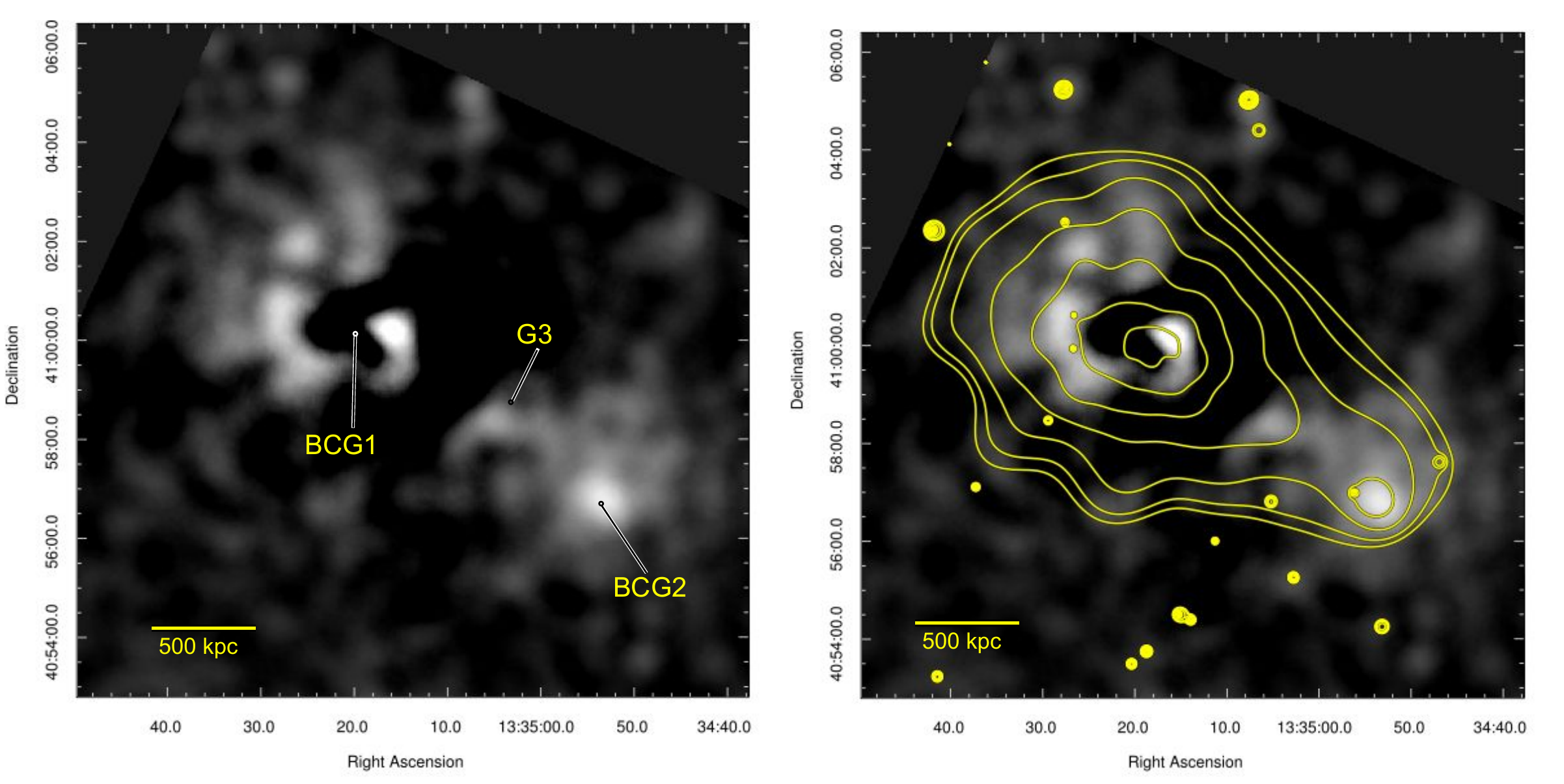}
\caption{Residual image of excess X-ray emission (left) overlaid with contours of adaptively smoothed ICM emission from Figure \ref{fig:csmooth3Mpc}  (right).  The image was created by subtracting a 2D elliptical $\beta$-model from an unbinned image from which point sources had been excluded.  The residuals were then smoothed with a radius of $\sigma$ = 16.3$"$.  The BCG2 subcluster was masked out in the fitting process, but unmasked for $\beta$-model subtraction. The three brightest galaxies in the system (BCG1, BCG2, and G3) are labeled. }
\label{fig:excess}
\end{center}
\end{figure}

We subtracted the 2-D $\beta$-model from an unsmoothed image of cluster emission. Smoothing the residuals ($\sigma = 16.3"$, 60 kpc) reveals a number of excess surface brightness features present in the gas (Figure \ref{fig:excess}).  A prominent spiral-type structure winds clockwise, outward from the western emission peak to a radius of $\sim$ 850 kpc northeast of the core. The structure is comparable in shape to sloshing spirals seen in many SCC clusters \citep[e.g.][]{Paterno13,Blanton11,Lagana10}. To the southwest of the core, the BCG2 subcluster is visible with an elongated feature pointing back toward the primary system.  This elongated feature is coincident with G3, the third-ranked galaxy in the system. Each excess feature is detected above the underlying 2-D elliptical model with greater than 3-$\sigma$ confidence.

\begin{figure}
\begin{center}
\hspace*{-1cm}  
\includegraphics[width=200mm]{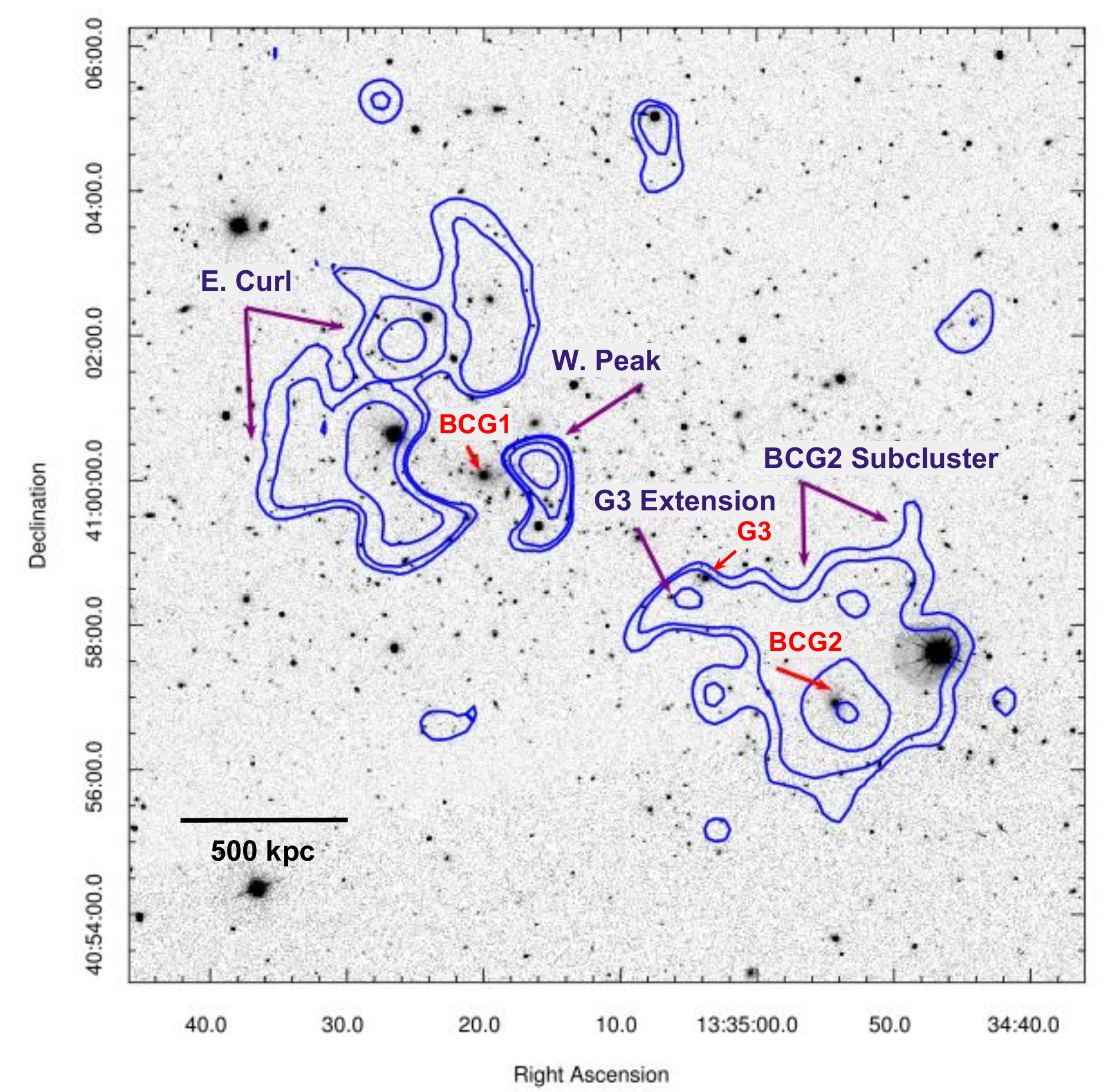}
\caption{Contours of residual emission overlaid on an SDSS r-band image of Abell 1763.  Locations of ICM excesses are labeled.  The three brightest galaxies in the system (BCG1, BCG2, and G3) are identified in red.}
\label{fig:excessSDSS}
\end{center}
\end{figure}

The excess features along with the three brightest galaxies in the system (BCG1, BCG2, and G3) are labeled in Figure \ref{fig:excessSDSS}.   It shows that the spiral structure (W. Peak + E. Curl) occupies a large portion of the projected area of the primary cluster.  The western peak, offset to the west of BCG1 by roughly 150 kpc, extends south and eastward from its brightest point.  Continuing eastward, beyond a decrement in residual emission, a large broad feature curls up in a clockwise direction, giving the appearance of a coherent structure spiraling outward from the western peak.

The BCG2 subcluster can be seen in Figure \ref{fig:excess} and Figure \ref{fig:excessSDSS} in the southwestern quadrant of each image.  Except for the emission comprising the G3 extension, the gas distribution is relatively symmetrical, with the second brightest galaxy in the system (BCG2) lying at its center.

\subsubsection{Gaussian Gradient Magnitude (GGM) Filter}
To further probe the presence of substructure within the core, we applied the Gaussian gradient magnitude (GGM) filter to the data. First introduced to X-ray analysis in \citet{Sanders16a}  \citep[see also][]{Sanders16b} GGM filtering assumes Gaussian derivatives to calculate gradients in the two dimensional data. The resulting image shows bright and dark regions corresponding to steep and shallow gradients, respectively. GGM analysis allows for values of the Gaussian $\sigma$ to be adjusted (to measure gradients over larger or smaller radii) depending on the quality of the data and the size of the features of interest.  Composite images useful for analysis can then be created by combining GGM-filtered images produced from a range of $\sigma$ values, provided they are weighted appropriately.  Following the radial weighting scheme outlined in \citet{Sanders16b}, near the core we weighted the GGM-filtered images heavily with small $\sigma$, while those with larger $\sigma$ are weighted to display more prominently at larger radii. Smoothing radii of $\sigma$ = 5, 10, 20, and 30 pixels are used.  

\begin{figure}
\begin{center}
\hspace*{-1cm}  
\includegraphics[width=200mm]{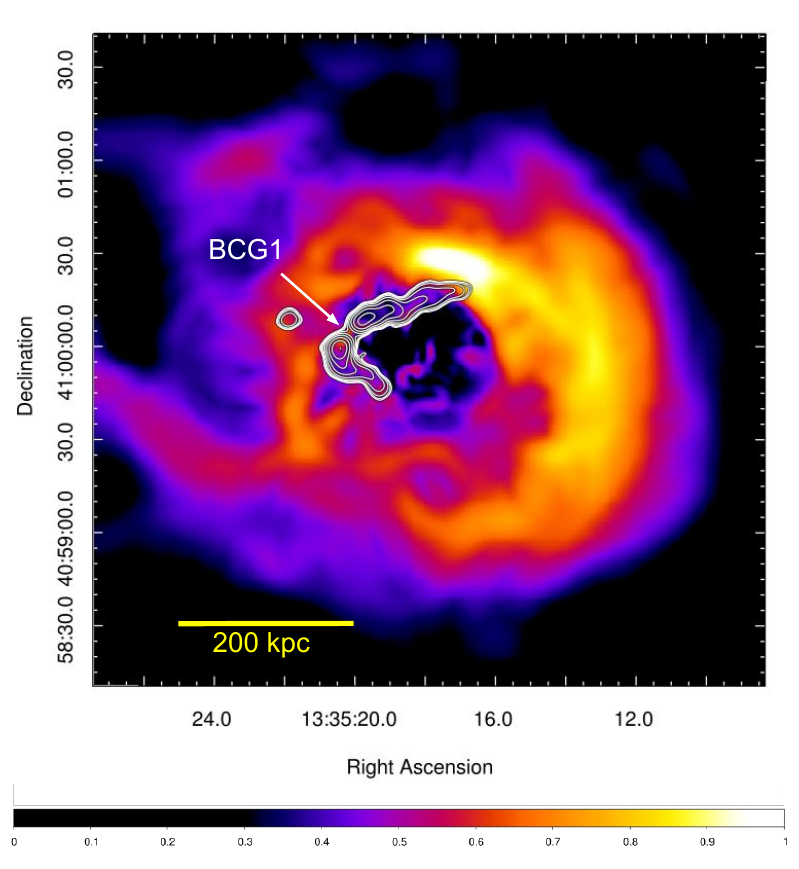}
\caption{ Gaussian gradient magnitude filtered image of the cluster core. The image combines smoothing radii of 5, 10, 20, and 30 pixels.  A clear arcing spiral structure can be seen extending clockwise from a location just north of the western peak. In addition to the spiral, an edge is present $\sim$ 130 kpc  to the northeast of BCG1.}
\label{fig:GGM}
\end{center}
\end{figure}

\begin{figure}
\begin{center}
\hspace*{-1cm}  
\includegraphics[width=200mm]{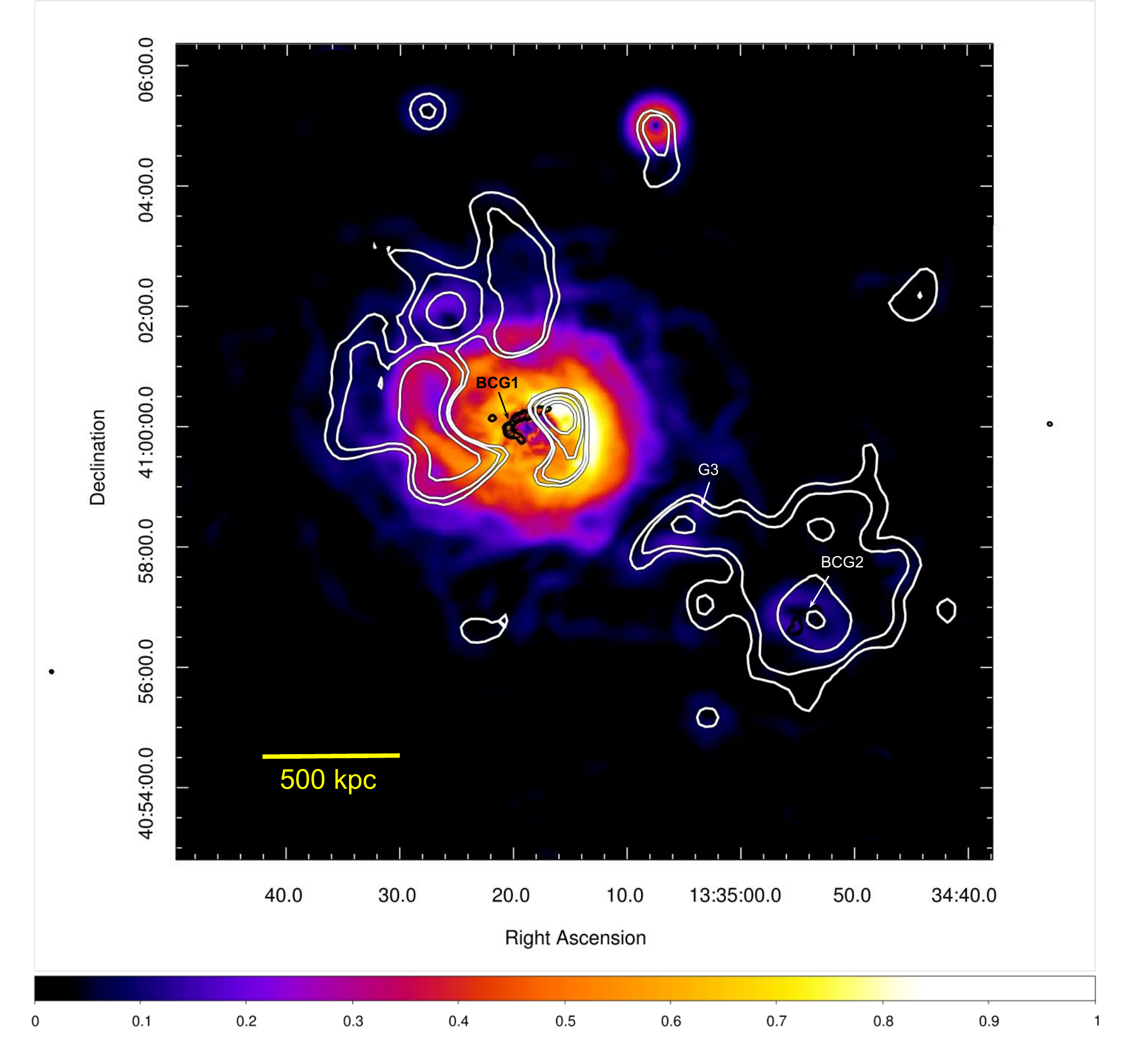}
\caption{Same GGM-filtered image as in Figure \ref{fig:GGM} shown covering an area 3 Mpc in size.  The stretch is adjusted to display lower intensity features at larger radii. Contours of residual emission (Figure \ref{fig:excess}) are overlaid in white. The outer edge of the spiral's western component is traced by a steep gradient in ICM emission, consistent with the feature having formed from a sloshing SCC.}
\label{fig:GGM_spiral}
\end{center}
\end{figure}

The resulting GGM image is presented in Figures \ref{fig:GGM} \& \ref{fig:GGM_spiral}.    In Figure \ref{fig:GGM} a clear spiral is visible.  It extends clockwise from just north of the western peak. As shown in Figure \ref{fig:GGM_spiral}, this steep gradient feature traces the outer edge of the western portion of the ICM spiral.  A steep gradient in emission along this outer western edge is consistent with the ICM spiral structure having been produced by the outward propagation of cold fronts launched by a sloshing SCC.  The eastern portion of the spiral (eastern curl) is similarly associated with a steep gradient feature (though of lesser prominence). These results suggest that at the onset of core sloshing and spiral formation, an SCC was present in the core of Abell 1763.

%%%%%%%%%%%%%%%
%%%%%%%%%%%%%%%
\section{SPECTRAL ANALYSIS OF CLUSTER EMISSION} \label{sec:spectral}

\subsection{Total Spectrum of Primary Cluster and BCG2 Subcluster}\label{sec:globspec}
A spectrum was extracted from a region of radius 1 Mpc (272$"$) around the center of the cluster using the \texttt{specextract} script in CIAO.  All X-ray point sources identified using \emph{wavdetect} were excluded.  The spectrum was binned such that each energy bin contained a minimum of 20 counts after background subtraction.  The energy range was restricted to 0.7-7.0 keV. Using \emph{XSPEC version: 12.8.2} we fitted an absorbed \texttt{APEC} model to the spectrum. We adopt the solar abundance table from \citet{AG89}.  Temperature, abundance, and normalization were free parameters.  The absorption parameter was fixed at the Galactic value of $N_H = 0.92 \times 10^{20}$ cm$^{-2}$ \citep{DL90}.  Determined from $\sim$ 17,000 source counts, we found the average global temperature to be 8.09$_{-0.34}^{+0.35}$ keV with an abundance of 0.28$_{-0.06}^{+0.07}$ Z$_{\odot}$.  The fit was good with a $\chi^2$ of 336.01 for 306 degrees of freedom, resulting in a reduced $\chi^2$ of 1.1. A spectrum was extracted from a region of radius 400 kpc (109$"$) around the center of BCG2. Following the fitting method described above, from $\sim$ 1000 source counts we determined the temperature of the BCG2 subcluster gas to be 4.00$_{-0.44}^{+0.65}$ keV. The  abundance was fixed at the standard value of 0.3 Z$_{\odot}$. Similar to the results presented in \citet{Zhu16}, the temperature of the cluster as a function of radius was found to be flat out to r$_{500}$.  

\subsection{Spectral Maps}

\begin{figure}
\begin{center}
\hspace*{-1.2cm}  
\includegraphics[width=200mm]{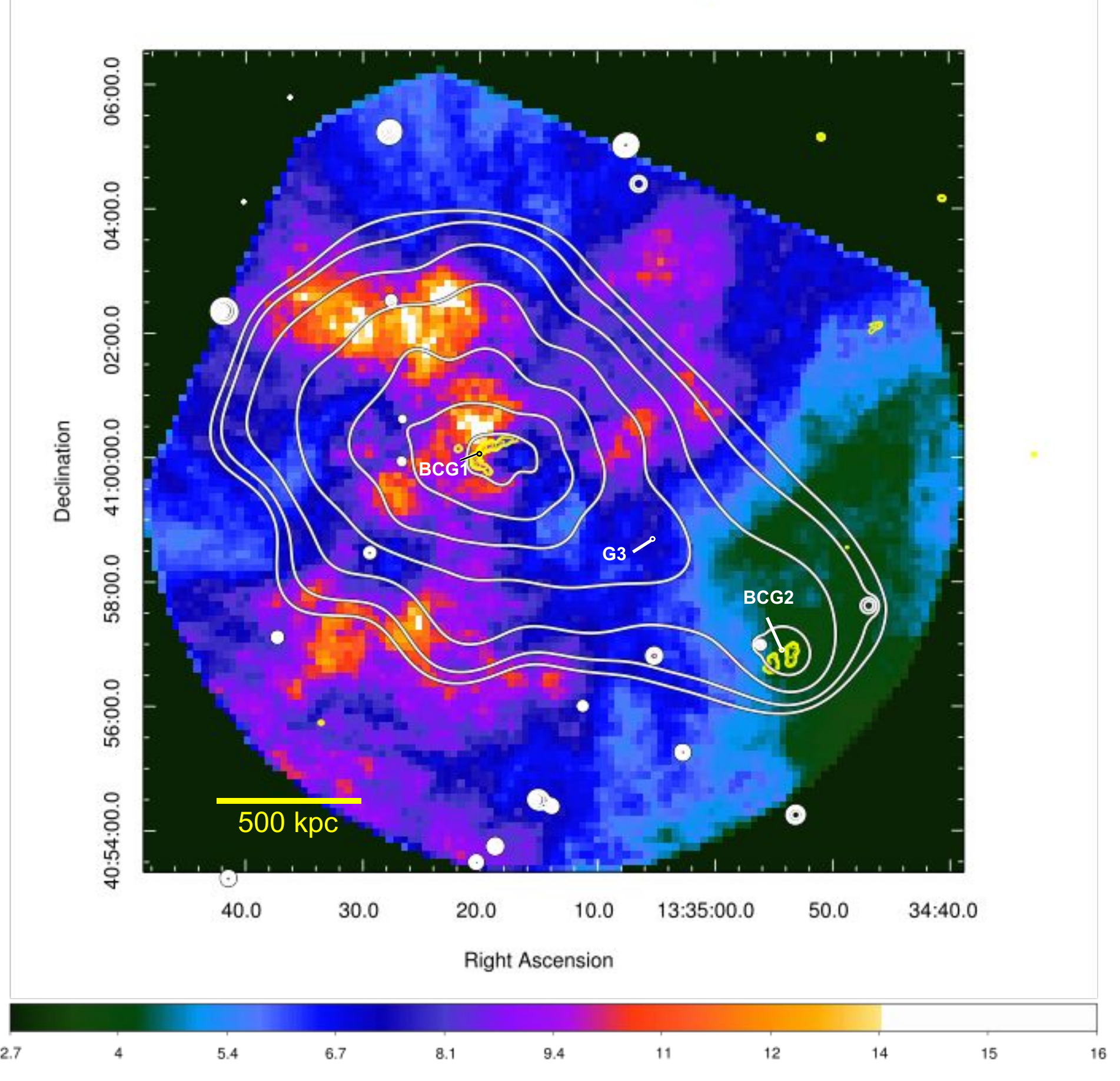}
\caption{Temperature map overlaid with contours of X-ray emission (white) and 1.4 GHz radio emission (yellow). Units are in keV.  Regions of high temperature are apparent to the northeast of the cluster core. The BCG2 subcluster is coincident with cooler gas.  The pointed spectral analysis of the northeastern high-temperature features is shown in Figure \ref{fig:SBpress}}.
\label{fig:tempMAP}
\end{center}
\end{figure}

\begin{figure}
\begin{center}
\hspace*{-1.2cm}  
\includegraphics[width=200mm]{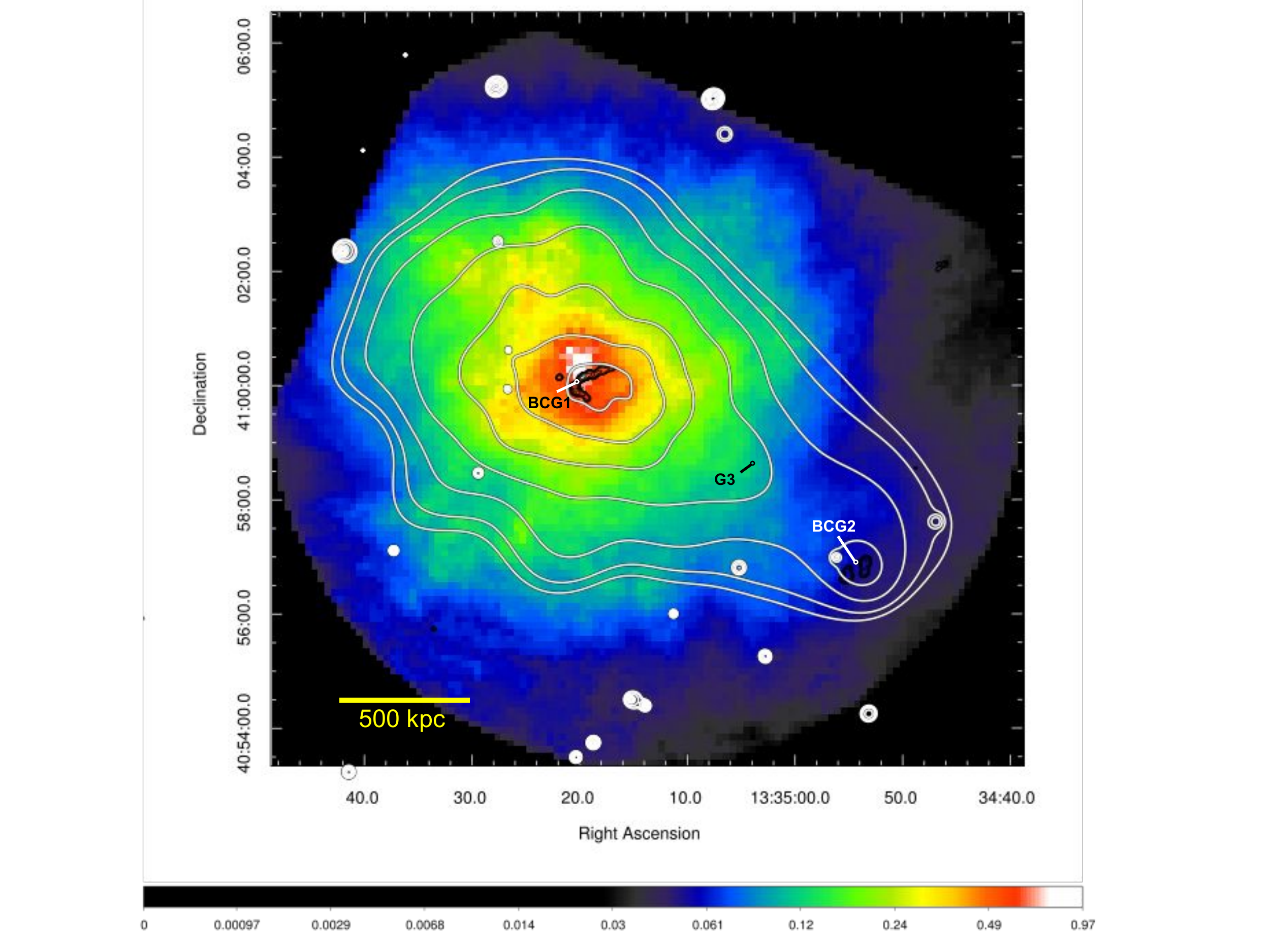}
\caption{Pressure map overlaid with contours of X-ray emission (white) and 1.4 GHz radio emission (black).  Relative values of pressure (unitless) are shown.  The area of highest pressure can be seen in a region to the northeast of the X-ray centroid and the central WAT source (hosted by BCG1).}
\label{fig:pressMAP}
\end{center}
\end{figure}

\begin{figure}
\begin{center}
\hspace*{-1.2cm}  
\includegraphics[width=200mm]{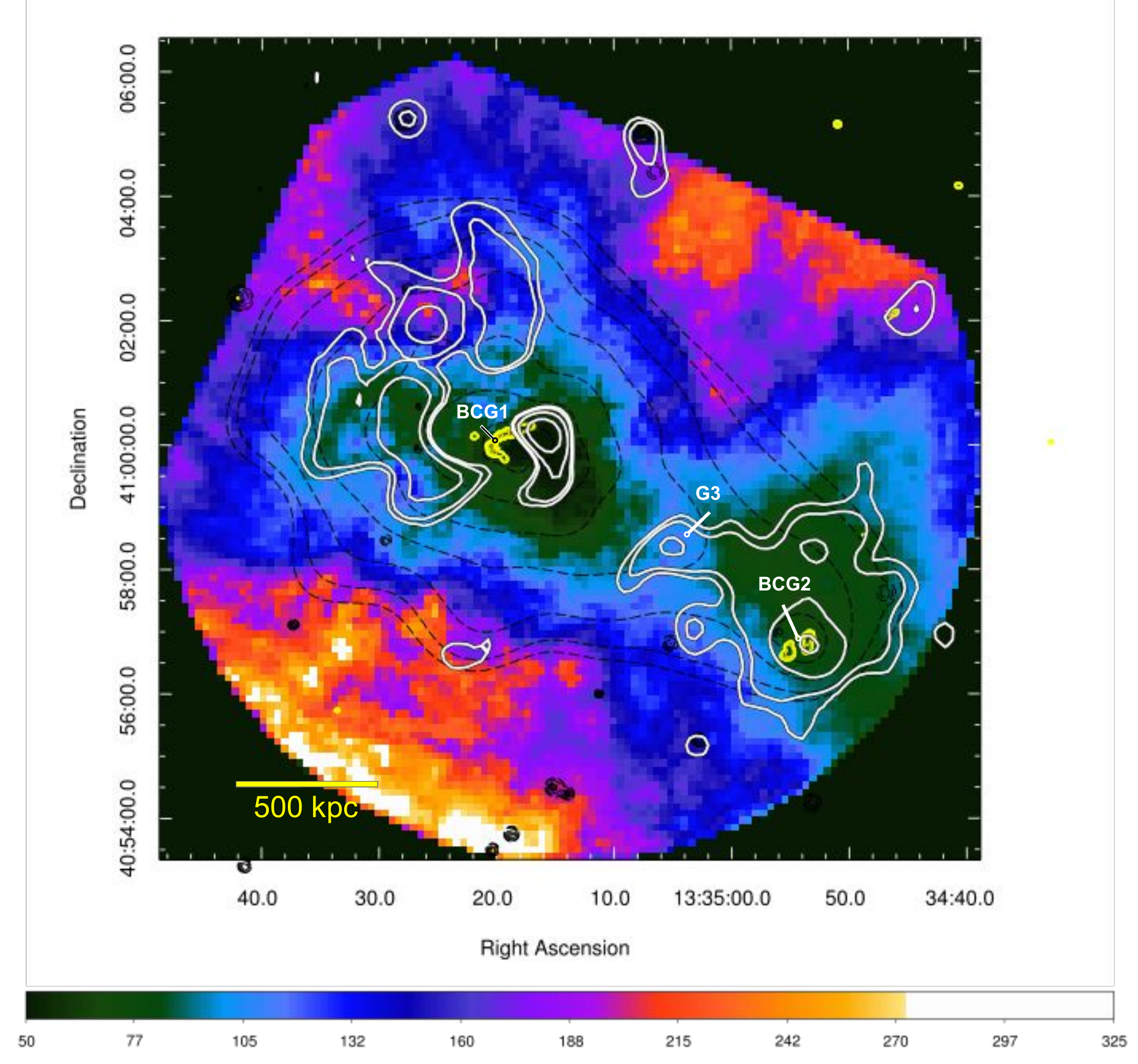}
\caption{Entropy map overlaid with contours of excess emission (white) and 1.4 GHz radio data (yellow).  Relative values of entropy (unitless) are shown. Regions of lower entropy appear coincident with the western emission peak, the lower eastern curl, and the BCG2 subcluster.}
\label{fig:entropMAP}
\end{center}
\end{figure}

Temperature, pressure, and entropy maps were created using the method described in \citet{Randall08,Randall09a}. The maps (Figures \ref{fig:tempMAP} - \ref{fig:entropMAP}) cover a circular region of radius 400$"$ (1.47 Mpc). The center of the region was chosen to ensure that both the main cluster and the BCG2 subcluster were within the map boundaries. Each map pixel value was determined by extracting a spectrum from the smallest surrounding circular region that contained 1000 total source counts in the 0.7-7.0 keV band. Each spectrum was fitted with an absorbed \texttt{APEC} model.  Galactic absorption was fixed at the value in the global fits. Abundance was allowed to be free\footnote{While abundance was a free parameter in the 2D temperature map fitting process, the low number of counts resulted in poorly constrained values. Due to the lack of discernible structure in the abundance map produced in the fit, and because large uncertainties are associated with the fitted values, we chose not to include it in this paper.}. Values shown in the 2D maps have uncertainties ranging from 25\% near the core to greater than 50\% at a radius of 1 Mpc. Smearing effects are also present due to the often large radii of the adaptively sized spectral extraction regions. Therefore, to gauge the significance of thermal features apparent in the maps, we perform pointed spectral extraction and analysis on regions of interest.  Profiles of surface brightness, temperature, density, pressure, entropy, and cooling time are discussed in Section \ref{sec:radial} and shown in Figures \ref{fig:SBpress} and \ref{fig:Ktcool}.

\subsubsection{Temperature Map}
The temperature map is shown in Figure \ref{fig:tempMAP}. The BCG2 subcluster is identified as the region of $\sim$ 4 keV gas to the southwest of the primary, consistent with the temperature determined from the 400 kpc region described above.  Immediately to the northeast of the primary cluster core, a region of high-temperature gas is apparent, extending to a distance $\sim$150 kpc from BCG1. A second larger region of enhanced temperature lies beyond this, along the same northeastern axis, extending to a radius of $\sim$ 1 Mpc. A pointed analysis of this northeastern sector of the cluster is performed in Section \ref{sec:NNE}. There does not appear to be notable temperature structure in the region between the primary system and BCG2 subcluster.

\subsubsection{Pressure Map}\label{sec:spectralpress}
The pressure map in Figure \ref{fig:pressMAP} appears to show an axisymmetric pressure distribution.  However, the peak of the pressure distribution is not coincident with the peak of the X-ray emission.  Instead, the region of highest pressure lies just to the northeast of BCG1, extending to a radius of $\sim$150 kpc. The map shows gas of enhanced pressure extending outward further beyond this feature into the northeastern sector of the cluster. Radial profiles of these northeastern regions are presented in Section \ref{sec:NNE}.

\subsubsection{Entropy Map}

The entropy map is shown in Figure \ref{fig:entropMAP}.  Contours of excess emission from Figure \ref{fig:excess} are overlaid.  Two prominent regions of lower entropy can be seen coincident with the primary cluster and BCG2 subcluster.  The western peak and its southern extension trace the regions of lowest entropy gas on the map. On a larger scale, intermediate-entropy gas appears to extend in both directions along the NE-SW elongation axis of the system. Profiles of entropy and cooling time are presented in Section \ref{sec:WSW} for a wedge region centered on the western peak, extending toward the southwest, beyond the BCG2 subcluster.  

\subsection{Radial Profiles}\label{sec:radial}
To further investigate interesting regions identified in the spectral maps (Figures \ref{fig:tempMAP} - \ref{fig:entropMAP}), we have created surface brightness profiles and extracted spectra from wedges of partial concentric annuli (see insets in Figures \ref{fig:SBpress} and \ref{fig:Ktcool}). Spectral extraction region sizes were chosen to include at least 500 source counts in the 0.7-7.0 keV energy range.  Spectra were grouped to 20 counts per bin and fitted with an absorbed APEC model, with abundance fixed at the globally fitted value of Z = 0.28. Best-fit parameters were determined by minimizing the C-statistics. When determining profiles of density and pressure, surface brightness was deprojected assuming spherical shells of constant emissivity and temperature. Profiles are plotted with 1-$\sigma$ error bars.    
   
\subsubsection{Northeastern Shock-like Feature}\label{sec:NNE}
A region of enhanced temperature and pressure is apparent in the spectral maps (Figures \ref{fig:tempMAP} and \ref{fig:pressMAP}) extending $\sim$ 150 kpc northeast of BCG1.  To further probe this sector of the cluster, we created a surface brightness profile and extracted spectra from concentric partial annuli forming an 80$^{o}$ wedge centered on the eastern peak and extending out to a radius of $\sim$ 1 Mpc.  

\begin{figure}
\begin{center}
\hspace*{-1.2cm}  
\includegraphics[width=200mm]{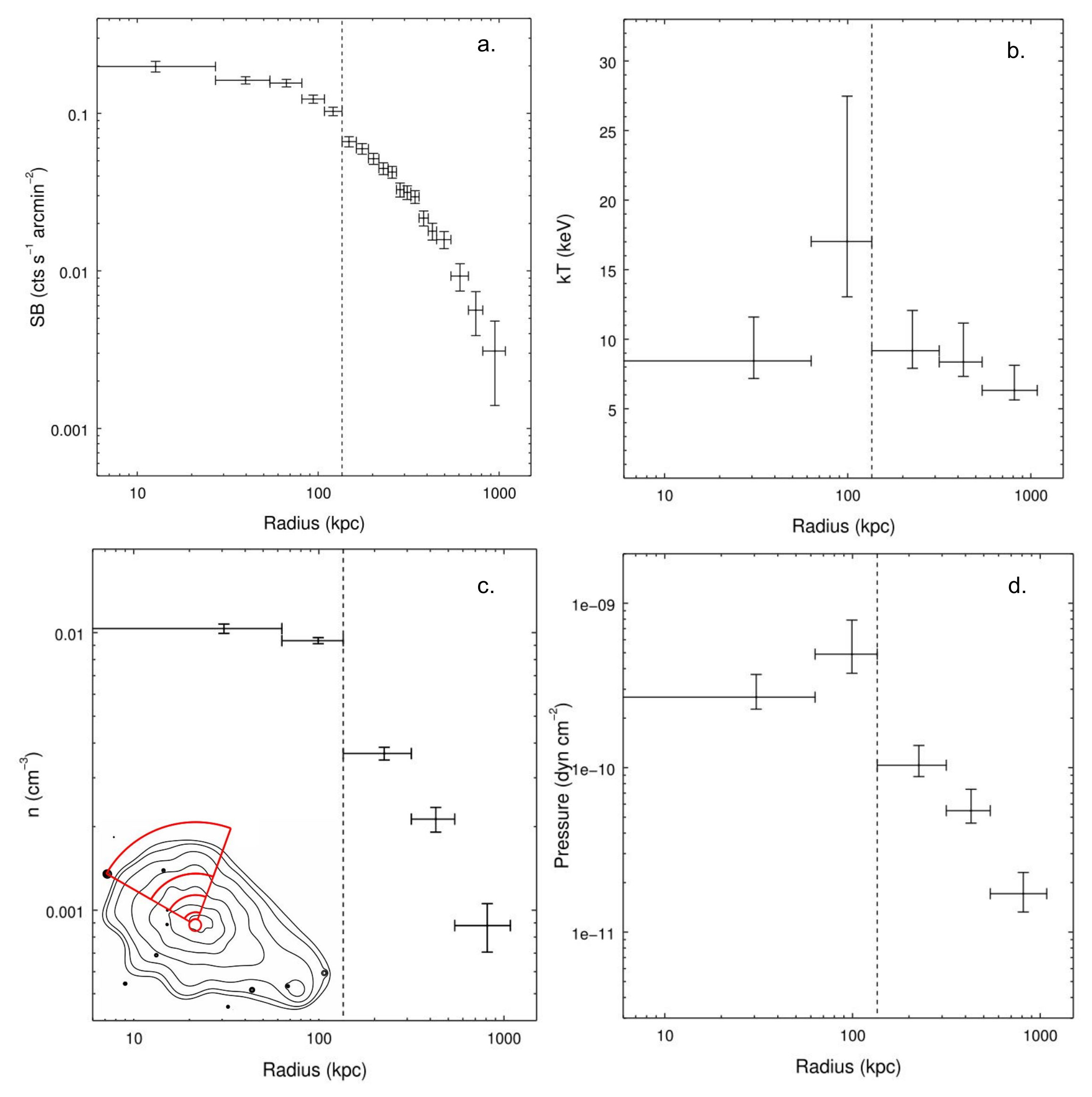}
\caption{(a) Surface brightness profile of a wedge of 80$^{\circ}$ partial annuli centered on the eastern emission peak, extending to a radius of $\sim$ 1 Mpc (see inset). A surface brightness edge at a radius of 135 kpc is detected at 3$\sigma$ significance. It is marked as a dashed line in all plots. (b) Temperature profile of the 80$^{\circ}$ wedge. A temperature difference is detected across the r=135 kpc interface at 1.6$\sigma$ significance. (c) Density profiles display a drop across the same interface. (d) Pressure profile is consistent with the shock interpretation.}
\label{fig:SBpress}
\end{center}
\end{figure}

The surface brightness profile shown in Figure \ref{fig:SBpress}a displays a significant discontinuity at a radius of 135 kpc. Surface brightness values on either side of the dashed line differ at the 4.6$\sigma$ confidence level. A 1-D beta-model (\emph{beta1d}) was fitted to the profile excluding points interior to the SB edge. Comparing the surface brightness of the region just interior to the edge with the model value at the same radius, we find a difference at a significance of 3.75$\sigma$.

The temperature profile of the 80$^{\circ}$ wedge is shown in Figure \ref{fig:SBpress}b. The temperature of the region corresponding to the enhanced temperature feature in the map (immediately to the northeast of BCG1) is determined to be 17.0$^{+10.7}_{-4.08}$ keV.  The neighboring region, external to the r=135 kpc (R$_{135}$) surface brightness edge, has a temperature of 9.17$^{+2.96}_{-1.29}$ keV. This corresponds to a temperature difference across the interface detected at 1.6$\sigma$ significance. The factor by which temperature drops across the edge is determined to be f$_{T}$ = 1.85$^{+1.21}_{-0.74}$. The high-temperature region lies interior to the R$_{135}$ interface, suggesting the surface brightness edge is due to a shock rather than subsonic motion of a cold front \citep{MV07}. Therefore, we can use Rankine-Hugoniot jump conditions to gain an estimate for shock velocity. Numerically, we determine a Mach number of $\mathcal{M}$ = 1.81$^{+0.84}_{-0.70}$, corresponding to a 1-$\sigma$ velocity range of 1576 km s$^{-1}$ $<$ v$_s$ $<$ 3763 km s$^{-1}$ (c$_s$ = 1420 km s$^{-1}$ for an 8 keV cluster). Profiles of density and pressure (Figures \ref{fig:SBpress}c and \ref{fig:SBpress}d) show prominent decreases across the R$_{135}$ interface that are consistent with the shock interpretation. As seen in the off-axis merger simulations of \citet{Zuhone11}, a shock is expected to develop during disruptive core sloshing as a cold front is launched into the high-velocity ICM counter flow induced by the infalling system (see Section \ref{sec:simdisc}).
%%%%%%%%%%%%%%%%%%%%%%%%%%%%%%%%%%
\subsubsection{Western Peak and and BCG2 Subcluster}\label{sec:WSW} 
\begin{figure}
\begin{center}
\hspace*{-1.2cm}  
\includegraphics[width=170mm]{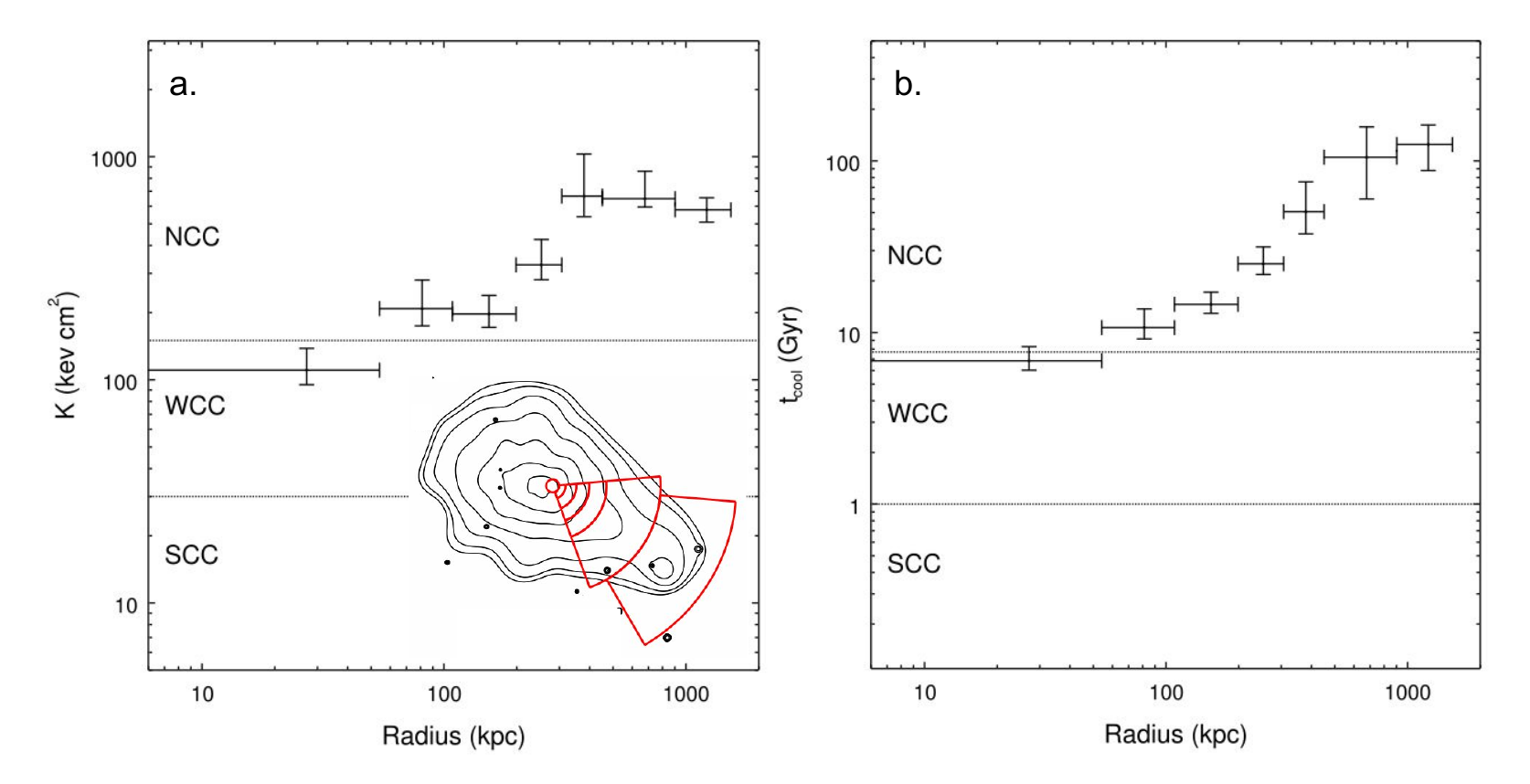}
\caption{(a) Profile of entropy and (b) cooling time for partial circular annuli centered on the western emission peak, extending beyond the BCG2 subcluster to a radius of $\sim$1.5 Mpc (see inset). The western emission peak is found to have the lowest entropy and cooling time in the cluster (K = 111$_{-15}^{+28}$ keV cm$^2$, t$_{cool}$ = 6.84$_{-0.80}^{+1.43}$ Gyr). Dashed lines indicate break values used to define SCC, WCC, and NCC core states \citep{Hudson10}. Despite the presence of a large gas-sloshing spiral, the cluster's core properties fall far above upper boundaries (in K$_0$ and t$_{cool}$) used to define the SCC class.}  
\label{fig:Ktcool}
\end{center}
\end{figure}

To probe the entropy of the western peak and region between the primary cluster and BCG2 subcluster, profiles were created for a wedge of partial annuli subtending an angle of 75$^{\circ}$ (Figure \ref{fig:Ktcool}). The wedges are centered on the western peak and extend beyond the BCG2 subcluster to a radius of $\sim$ 1.5 Mpc. The temperature of the western peak is determined to be 5.93$_{-0.8}^{+1.5}$ keV. Though not shown, the temperature profile is flat out to 1 Mpc where it decreases to a value near that quoted for the BCG2 subcluster in Section \ref{sec:globspec}. The entropy of the western peak (Figure \ref{fig:Ktcool}a) is determined to be K$ = 111_{-15}^{+28}$ keV cm$^{2}$, which increases with increasing radius before decreasing again at the radius of the BCG2 subcluster. Dotted lines show the central entropy break values demarcating SCC, WCC, and NCC clusters in \citet{Hudson10}.  From the analysis discussed in Section \ref{sec:NNE}, the entropy of the eastern peak was determined to be K$ = 177_{-26}^{+66}$ keV cm$^{2}$. This indicates that the western peak is the region of lowest entropy within Abell 1763. That the lowest entropy gas of the system is found in a region 150 kpc to the west of BCG1 is evidence that significant gas displacement has taken place. The intermediate entropy and location of the western peak are consistent with it being the cluster's former SCC, disrupted during the spiral formation process.

The cooling time was calculated in the same wedge extending to the southwest from which entropy was determined (Figure \ref{fig:Ktcool}b). Dotted lines are included to show the central cooling time break values used for classifying SCC, WCC, and NCC clusters \citet{Hudson10}.  The cooling time of the western peak was determined to be $t_{cool} = 6.84_{-0.80}^{+1.43}$ Gyr which is in the regime of WCCs, well above the SCC threshold of 1 Gyr. The cooling time of the eastern peak was determined to be $t_{cool} = 8.23_{-1.2}^{+2.7}$ Gyr; a value more consistent with those seen in the cores of NCC clusters.

%%%%%%%%%%%%%%%
%%%%%%%%%%%%%%%
\section{GALAXY VELOCITIES}\label{sec:velocities}

\begin{figure}
\begin{center}
\hspace*{-0.5cm}  
\includegraphics[width=170mm]{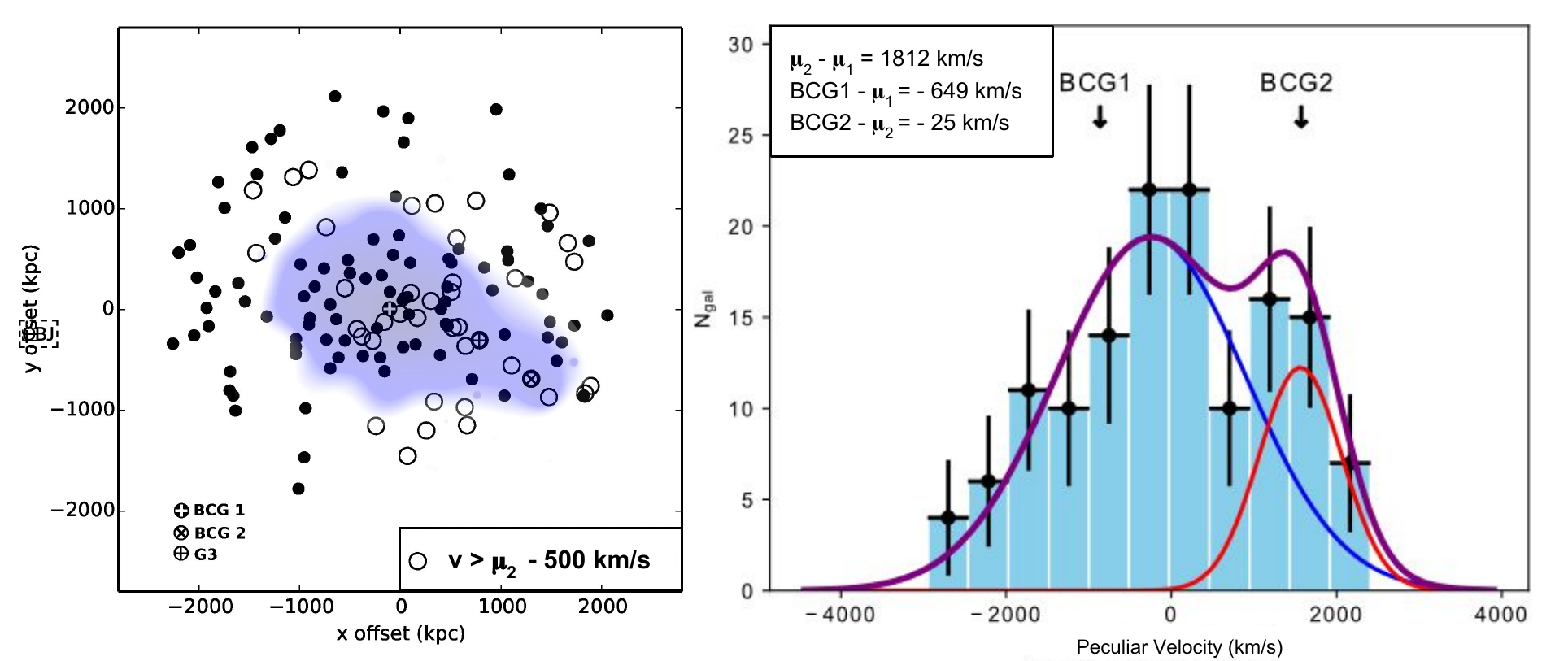}
\caption{Left panel: Distribution on the sky of 137 member galaxies of Abell 1763 with measured redshifts in \citet{Rines13} and SDSS DR-13. Filled and empty circles represent galaxies with peculiar velocities below and above v$_{div}$, respectively (where v$_{div}$ = $\mu_2$ - $\sigma_2$,  $\sigma_2$ = 500 km s$^{-1}$).  Though $\sigma_2$ is an estimated (non-fitted) parameter, the division it produces highlights the overabundance of high peculiar velocity galaxies in the western sector of the cluster.  Right panel: Distribution in velocity space of cluster members.  Bin widths are 500 km s$^{-1}$. A two-component Gaussian mixture model was fitted to the velocity distribution. Blue, red, and purple curves show the primary Gaussian, secondary Gaussian and composite respectively.  The velocity offset between the two components is v = 1812$_{-349}^{+399}$ km s$^{-1}$. BCG1 is offset from the mean of the primary cluster by -649$_{-276}^{+239}$ km s$^{-1}$, while BCG2 is offset from the secondary mean ($\mu_2$) by -25$_{-255}^{+285}$ km s$^{-1}$.}
\label{fig:HISTds}
\end{center}
\end{figure}

We conducted a dynamical analysis of the cluster using a total of 137 galaxy redshifts (Figure \ref{fig:HISTds}). The sample was compiled from those published in \citet{Rines13} (115 galaxies) and others made available in SDSS-DR13 (22 galaxies, including  BCG1 and BCG2).  The distribution on the sky of the 137 galaxies included in the analysis is shown in Figure \ref{fig:HISTds} - left.  The mean redshift was found to be z = 0.231. We calculated the peculiar velocity for each galaxy in the frame of the cluster \citep[v$_{pec,i}$ =(v$_{mean}$ - v$_{i}$)/(1+z$_{mean}$),][]{Faber77}. After redshift correction, the velocity dispersion of the cluster was calculated to be $\sigma$ = 1287 $\pm$ {47} km s$^{-1}$, consistent with an 8 keV cluster based on the $\sigma$-T relation of \citet{Wu99}. A histogram of the 137 corrected peculiar velocities is shown in Figure \ref{fig:HISTds} - right. Velocities are binned at 500 km s$^{-1}$, and the uncertainty in galaxy counts for each bin is determined using the approximation for low count data: 1 + (N + 0.75)$^{1/2}$ \citep{Gehrels86}, where N is number of galaxies per bin. The location in velocity space of BCG1 and BCG2 is also shown.  

\subsection{Single-Gaussian Fit to Cluster Velocities}
 A single-component Gaussian model was fitted to the one-dimensional peculiar velocity data.  Amplitude (A), mean ($\mu$), and velocity dispersion ($\sigma$) were all free parameters.  The data was fitted using the \emph{least squares} algorithm in \emph{matplotlib} (Python v 2.7).  To determine parameter values and confidence intervals, 1000 simulated data sets were created using the individual uncertainty values of each bin to define the normal distribution from which the simulated data points were drawn.  The fitted values quoted are the median fitted values of the 1000 fits.  The range of fitted parameter values returned did not necessarily follow a normal distribution, therefore asymmetrical ``1-sigma" confidence intervals were drawn at the +/- 34\% mark from the median values.  Results, including confidence intervals, and reduced $\chi^2$ are shown in Table \ref{tab:vels}.  While the single-component Gaussian model is a relatively good fit to the data, the cluster displays strong evidence of a large-scale multicomponent interaction.

\subsection{Double-Gaussian Fit to Cluster Velocities}
The large-scale X-ray morphology of Abell 1763 and the 2440 km/s offset between BCG1 and BCG2 are consistent with an ongoing binary merger. To investigate whether a two-component merger is evident in dynamical data, a mixture model of two summed Gaussian distributions was fitted to the galaxy velocities. We allowed amplitudes (A$_1$,A$_2$) and means ($\mu_1$,$\mu_2$) of the two Gaussians to be free.  To probe the presence of the BCG2 subcluster and its effect on the determination of the primary cluster's mean, the velocity dispersion ($\sigma_1$) of the primary system was allowed to be free, while the velocity dispersion of the secondary ($\sigma_2$) was fixed at 500 km s$^{-1}$.  This was done to counter the tendency of the fitting routine to return a sharp peak (A$_2$ $>$ A$_1$ \& $\sigma_2$ $<$ 200 km s$^{-1}$) at the location of the secondary. The reduced $\chi^2$ of the double-Gaussian fit is slightly larger than that of the single-Gaussian.  However, in an exploration of bin widths, the fits consistently located the secondary system at the same position in velocity space as BCG2.  

The primary and secondary systems were identified in the fit as having a velocity offset of 1812$_{-349}^{+399}$ km s$^{-1}$.  The offset between BCG1 and $\mu_1$ was determined to be -649$_{-276}^{+239}$ km s$^{-1}$ while the offset between BCG2 and $\mu_2$ was -25$_{-255}^{+285}$ km s$^{-1}$.   The fitting of the secondary system in the vicinity of BCG2 (in velocity space) indicates that the BCG2 subcluster, apparent in the X-ray image, is moving with a high line-of-sight velocity relative to the primary. The high relative velocity between the two systems is likely related to the formation of the ICM merger features detailed in previous sections. The dynamical complexity of the core is pronounced, consistent with the ongoing disruption of an ICM core that was likely, until recently, in an SCC state.

\subsubsection{High Peculiar Velocity of BCG1}
In each of the three methods employed to establish the mean of the primary cluster velocity distribution, a high peculiar velocity is observed for BCG1.  The peculiar velocity for BCG1 is calculated to be v$_{pec}$ = -871 km s$^{-1}$, -1048 km s$^{-1}$, and -649 km s$^{-1}$ for the algebraic, single-Gaussian, and double-Gaussian methods, respectively.  The velocity offset was detected in the dynamical analysis of \citet{Fadda08} as well, with their sample of 100 galaxies. Such an offset is not typically observed in off-axis mergers. While merging clusters may be characterized by bimodal velocity distributions, accounting for the presence of two interacting systems, the peculiar velocities of cluster dominant galaxies should remain low when measured against the evolving mean of their parent clusters \citep[e.g.][]{Coziol09}. Recent binary merger simulations of J. ZuHone however, produce unexpectedly high peculiar velocities of cluster dominant galaxies at times near core crossing. This disturbed phase space of the cluster center likely results from the differential gravitational forces experienced by the DM cores versus the outer regions of the cluster. This pronounced dynamical complexity in the core of Abell 1763 is undoubtedly linked to the destruction of the SCC.

In Section \ref{sec:discG3} we discuss the third-ranked galaxy (G3) in the system which is likely the dominant galaxy in an additional infalling subcluster. It is located at an intermediate distance between BCG1 and BCG2 (distance to BCG1 $\sim$ 750 kpc). Its trajectory and velocity relative to the primary mean $\Delta$v $\sim$ 1500 km s$^{-1}$, allow for the conclusion that the system's ICM halo and associated DM structure might be contributing prominently to core sloshing and spiral formation. It may also be playing a substantial role in generating the large peculiar velocity of BCG1, imparting upon it significant gravitational acceleration in an already complex dynamical environment.

%%%%%%%%%%%%%%%
%%%%%%%%%%%%%%%
\section{DISCUSSION}\label{sec:disc}
\subsection{The Megaparsec-Scale Spiral}
Since the launch of the \emph{Chandra} and \emph{XMM-Newton} observatories, low-entropy spirals of excess emission have been identified in a multitude of galaxy clusters \citep[e.g.][]{Blanton11, Lagana10, Johnson10, Clarke04} and groups \citep{Randall09b}.  They are understood to form when an SCC-hosting system experiences high impact parameter infall of a subsystem. In such interactions, ICM ram pressure causes the SCC to become displaced from the collisionless DM peak.  By nature of the off-axis encounter, the infalling system imparts angular momentum to the core gas.  The SCC then oscillates about the bottom of the cluster's gravitational potential well in a ``sloshing" motion \citep[see][]{MV07}.  Sloshing causes the formation of multiple subsonic cold fronts, which over time expand and merge to form the spiral patterns hosted by so many seemingly relaxed systems.  Simulations show that throughout the formation process, relative velocity between the SCC and DM peak generally remains low  \citep[$\Delta$v $<$ 400 km s$^{-1}$][]{AM06} and the average SCC-DM peak offset is less than $\Delta$r $<$ 30 kpc \citep{Zuhone10}.  

\begin{figure}
\begin{center}
\hspace*{-0.5cm}  
\includegraphics[width=170mm]{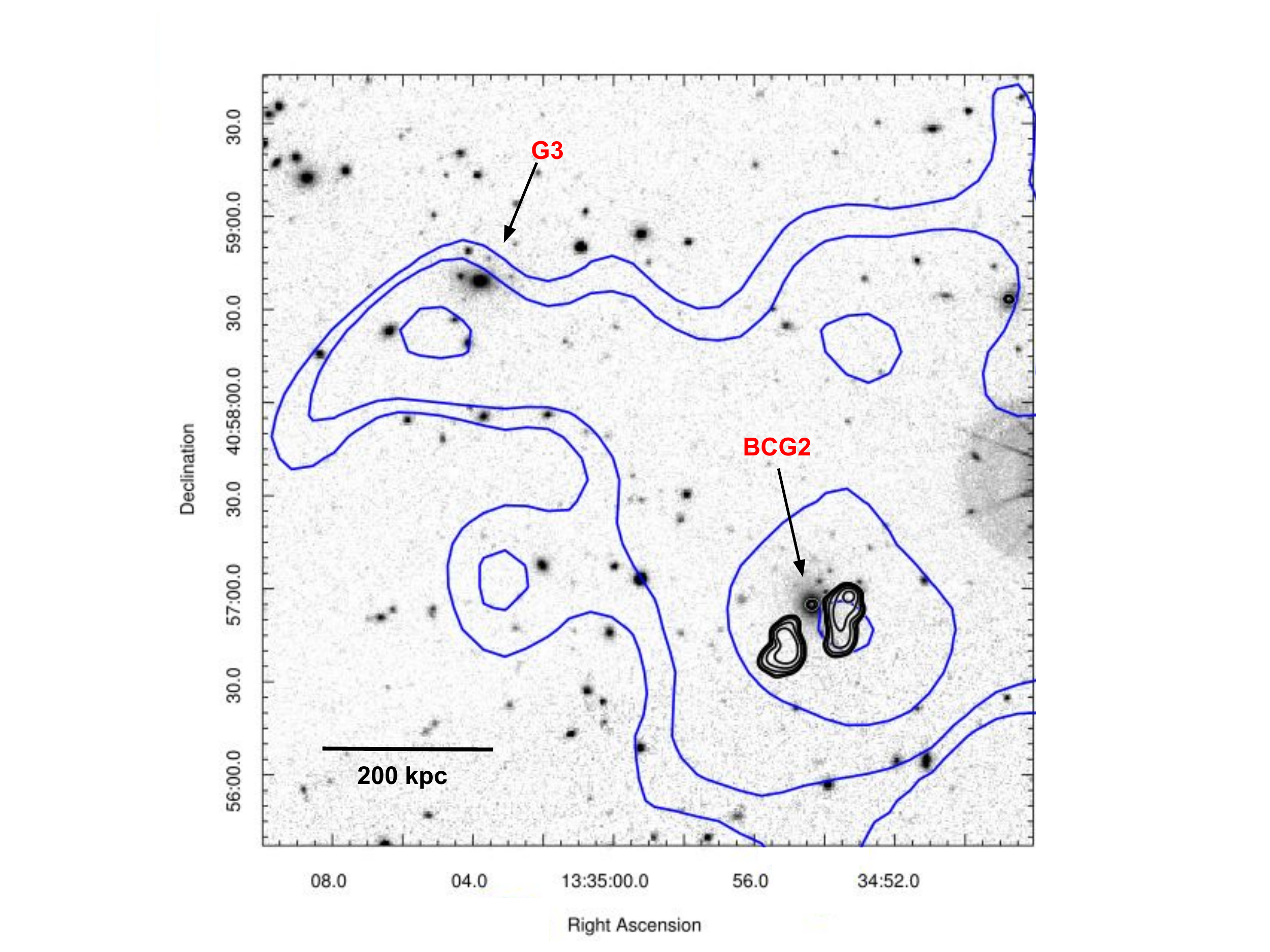}
\caption{ SDSS \emph{r-band} image of the southwestern region of the A1763 system.  Blue contours are residual X-ray emission, black contours are 1.4 GHz radio emission.  BCG2 lies at the center of the subcluster (bottom right) and is host to a bent WAT source.  The bending of the WAT lobes toward the south indicates the plane-of-the-sky component of motion is northward. The third-ranked galaxy (G3) is seen to the northeast of the BCG2 subcluster, roughly 50 kpc northwest of the brightest portion of the G3 extension. }
\label{fig:SCcon}
\end{center}
\end{figure}

\begin{figure}
\begin{center}
\hspace*{-0.5cm}  
\includegraphics[width=170mm]{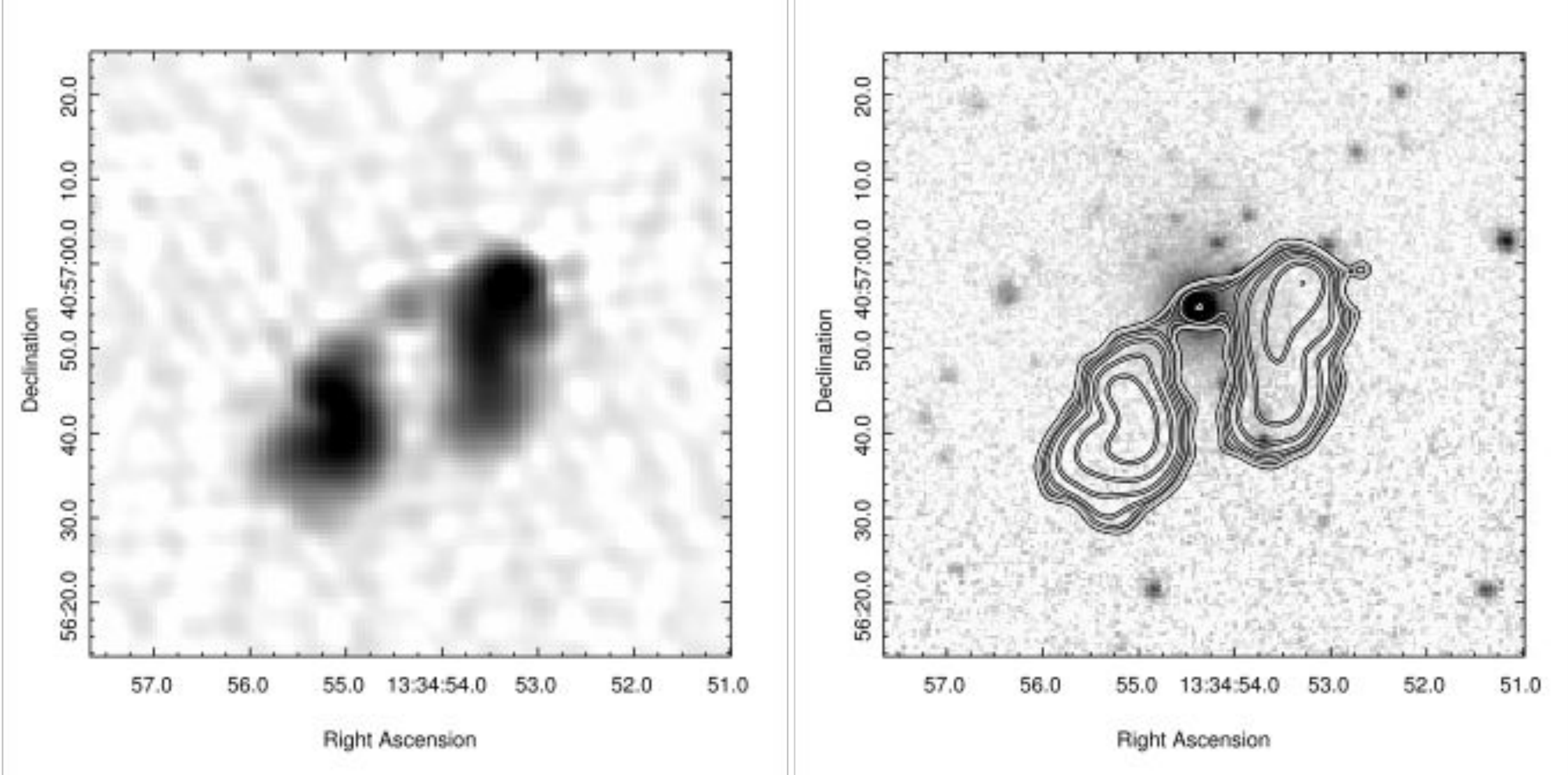}
\caption{VLA L-band image of the WAT associated with BCG2. The rms level for the 1.4 GHz image is 31 $\mu$Jy/beam with a beam of 3.6$\times$3.0 arcsec at a position angle of -69$^\circ$. The contours increase logarithmically from 10 times the rms level.}
\label{fig:WAT2}
\end{center}
\end{figure}

Based on simulations of off-axis mergers in SCC systems  \citep{AM06, Zuhone10, Zuhone11, Roediger11}, the winding direction of a spiral (when traced inward from a large radius), indicates the perturbing cluster's trajectory (clockwise vs. counterclockwise).  In Abell 1763, the position of the BCG2 subcluster with respect to the spiral makes it a promising candidate for having been the perturber.  This scenario requires that the BCG2 subcluster would have passed south of the primary system as it moved toward the west.  Following projected pericentric passage, it would have then swung toward the north. Such a trajectory is consistent with the bending angle of the WAT radio source hosted by BCG2 (Figure \ref{fig:SCcon}). The lobes of the WAT are bent toward the south (Figure \ref{fig:WAT2}), indicating a clear northward motion of the host galaxy relative to the surrounding ICM \citep{Gomez97}.  The counterclockwise trajectory of the BCG2 subcluster is consistent with its entrance to the Abell 1763 system via the intercluster filament \cite[see][]{Fadda08, Edwards10}, which extends toward the smaller cluster Abell 1770 (z=0.22), 12 Mpc to the northeast of Abell 1763.

\subsection{Cool Core Destruction in Conjunction with Spiral Formation}
Fundamental to the sloshing core model of spiral formation \citep{MV07} is the notion that SCCs should be found embedded in the centers of spiral-hosting systems.  This is understandable, as it is the oscillating SCC that produces the alternating cold fronts from which the spiral pattern forms. The infall events responsible for the majority of observed sloshing SCCs likely produce relative velocities between the SCC and DM peak that are insufficient to drive significant core disruption \citep{Zuhone10},    

A notable exception to the non-disruptive sloshing paradigm is Abell 2142.  Detailed in \citet{Rossetti13}, \citet{Owers11}, and \citet{Markevitch00}, the system displays evidence of opposing cold fronts but lacks an SCC (with central entropy K$_0$ = 68 keV cm$^2$). The cluster hosts two BCGs of comparable brightness separated by a projected distance of 183 kpc and high relative velocity ($\Delta$v=1800 km s$^{-1}$).  The high relative velocity of the BCGs near the two cold fronts led \citet{Markevitch00} to attribute the origin of these features to a head-on merger. However, in a wider field of view \emph{XMM} observation (with exposure of 55 ks), \citet{Rossetti13} identified a third excess at a distance of 1 Mpc from the core. They conclude that the three features are part of a coherent spiral structure, produced via sloshing of a formerly intact SCC.  The prospect that a large spiral would form in an environment of such dynamical activity to destroy an SCC, suggests that more violent processes are at play than in standard spiral formation scenarios.

Like Abell 2142, Abell 1763 does not possess an SCC. The core is instead filled with gas of intermediate entropy. The eastern peak's measured entropy (K = 177$_{-26}^{+66}$ keV cm$^{2}$) is higher than that of the western peak (K = 111$_{-15}^{+28}$ keV cm$^2$), along with the absence of cool gas in the core. This suggests that the majority of lower entropy gas once coincident with BCG1 has been displaced and heated.  

\subsubsection{Comparison to Off-axis Merger Simulations}\label{sec:simdisc}
\begin{figure}
\begin{center}
\hspace*{-0.5cm}  
\includegraphics[width=170mm]{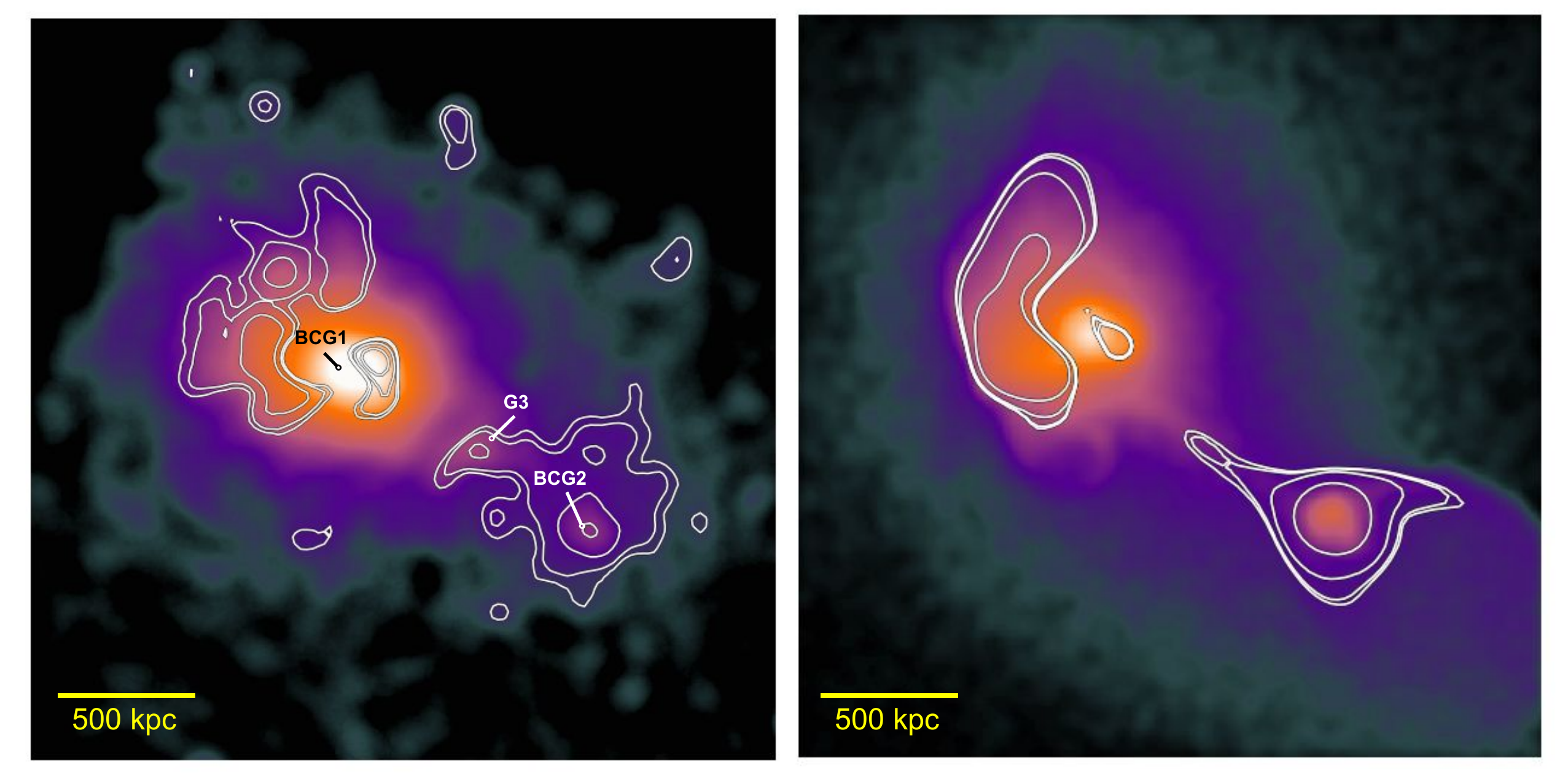}
\caption{Left panel:  \emph{Chandra} X-ray surface brightness distribution of Abell 1763 is overlaid with excess emission from Figure \ref{fig:excess}. Right panel: The X-ray surface brightness distribution of a 3:1 mass ratio, off-axis merger simulated in \citet{Zuhone11}.   Overlaid in white are contours of excess emission, determined following the method outlined in Section \ref{sec:spatial}. All four excess SB features present in Abell 1763 are also present in the simulated cluster.}
\label{fig:Zuhone}
\end{center}
\end{figure}

In an effort to better understand the ongoing infall event and probe possible merger configurations, we explore the merger simulations of \citet{Zuhone11}, available online in the Galaxy Cluster Merger Catalog\footnote{Galaxy Cluster Merger Catalog of \citet{Zuhone18} available at http://gcmc.hub.yt}. The binary merger simulations use an adaptive mesh-refinement grid-based code with Eulerian hydrodynamics. The underlying ICM model is one of inviscid, unmagnetized gas. We find that their run S6 (off-axis, 3:1 mass ratio, b=1 Mpc, v$_{t0}$= 1200 km s$^{-1}$), at an epoch of 1.35 Gyr since core crossing, displays a similar surface brightness distribution (z-axis projection) to that observed in Abell 1763 (Figure \ref{fig:Zuhone}).       

Apparent in the simulated Chandra observation, the subcluster lies $\sim$ 1.2 Mpc from the disturbed primary's center\footnote{It should be noted that the mass of the primary cluster in the simulation is 6 $\times 10^{14}$ M$_{\odot}$, roughly half that of the primary cluster in Abell 1763.}.  At this point in the simulation, the subcluster has reached its current position after passing south of the primary and is decelerating toward its point of minimum relative velocity before secondary infall.  Overlaid on the simulation image are contours of residual emission.  Residuals were determined in the same manner as in Section \ref{sec:spatial}.  All four surface brightness excesses in Abell 1763 are produced in the simulation.  Entropy maps of the simulated cluster show that during this epoch, the SCC disintegrates due to significant ram pressure induced in the core.  While an SCC originally existed, gas of intermediate entropy  now fills the central 200 kpc. The region of minimum entropy in that distribution is found in a clump extending 150 kpc from the core,  in the direction of the subcluster. This is similar to the western peak seen in Abell 1763. In the simulation, the eastern curl is produced as cool gas is launched from the core during pericentric passage in the form of an outward-propagating cold front.  \citet{Zuhone11} notes the presence of a  low-entropy bridge between the primary cluster and subcluster.   A similar feature is seen as the G3 extension to the east of the BCG2 subcluster in the Abell 1763 system (Figure \ref{fig:SCcon}). 

The agreement between the simulation and the surface brightness morphology of Abell 1763 is striking.  What is most interesting is that these features are reproduced in a dynamical environment that is vastly different to that of the Abell 1763 system.  At this epoch in the simulation, the subcluster is approaching its turn-around point, corresponding to a relative velocity to the primary system of zero.  The large velocity offset in Abell 1763 between the primary and BCG2 subcluster ($\Delta$v$\sim$1800 km s$^{-1}$) is a point of significant disagreement between observation and simulation. Nonetheless, the western peak, eastern curl, subcluster, and extension are all produced in the simulation. Additionally, at this epoch in their run, surface brightness discontinuities and regions of increased temperature develop north of the sloshing disrupted core. This could explain the shock-like feature observed to the northeast of BCG1 in Abell 1763.  However, upon analysis of simulated galaxy data at this epoch, we find that the high peculiar velocity of BCG1 is not reproduced.  

At about 1 Gyr after core passage, Kelvin-Helmholtz instabilities (KHIs) develop, which ultimately lead to a dissolution of the spiral and further disruption of the core. All mergers, in fact, in the parameter space exploration of \citet{Zuhone11} result in full transformation of SCCs into NCCs. Since these simulations consist of an unmagnetized inviscid fluid, the core and spiral have a greater vulnerability to the disruptive effects of KHIs than would be expected in a more realistic treatment of the ICM. When magnetic fields and gas viscosity are included in SCC sloshing simulations (albeit in more relaxed systems), the formation of KHIs can be suppressed, resulting in a longer-lived spiral structure and a more resilient core \citep[see][]{ZR16}. In Abell 1763, however, it is unlikely that KHI-suppressing phenomena such as magnetic fields and viscosity could be prominent enough to prevent full transformation of its core to an NCC state. Given its observed central entropy (K$_0$ $\sim$ 110 keV), heating and mixing of the core gas is well underway. This ongoing formation of an NCC in Abell 1763 by way of off-axis merger suggests that a broader set of dynamical conditions are capable of merger-driven destruction of SCCs than previously assumed (i.e. CC destruction is not exclusively relegated to head-on mergers).

\subsubsection{Timeline for Spiral Production}
Following the method outlined in \citet{Owers11} and \citet{Simionescu10}, we can obtain an estimate for the time since the onset of sloshing. This is done by calculating the time ($\tau$) required for two cold fronts to appear as opposing features across a cluster's center: 

\begin{equation}
\tau = \dfrac{\pi}{\omega_{k,in} - \omega_{k,out}}
\end{equation}

\noindent $\omega_{k,in}$ and $\omega_{k,outer}$ are the \emph{Keplerian} orbital frequencies at the radii of the inner and outer opposing cold fronts, respectively ($\omega_{k}$ = $\sqrt{2} \sigma$/R, where $\sigma$ = 1268 km s$^{-1}$). Given the relatively steep entropy profiles of galaxy clusters, $\omega_{k}$ is assumed to be a good approximation of the oscillation frequency of the superposed gravity waves that give rise to cold fronts \citep{Owers11,Simionescu10,Churazov03}. We calculate the time required to form the 850 kpc edge and its opposing cold front (250 kpc to the south\textbf{west} of the core) to be $\tau$ = 0.61 Gyr. The estimated age of the Abell 1763 spiral is significantly less than that of the simulated spiral ($\tau$ = 1.35 Gyr). The difference may be attributed, in part, to the lower mass of the simulated cluster. 

As mentioned above, the simulations show fluid instabilities become prominent within the spiral roughly $\sim$ 1 Gyr following pericentric passage of the subcluster.  Given the t $<$ 1 Gyr estimated time since onset of spiral formation, such instabilities may not have had time to grow to an observable scale in the system.  However, given the shallowness of the observation, it is unlikely that such features would be detectable if they were present.

\subsection{The G3 Extension}\label{sec:discG3}
\begin{figure}
\begin{center}
\hspace*{-0.5cm}  
\includegraphics[width=170mm]{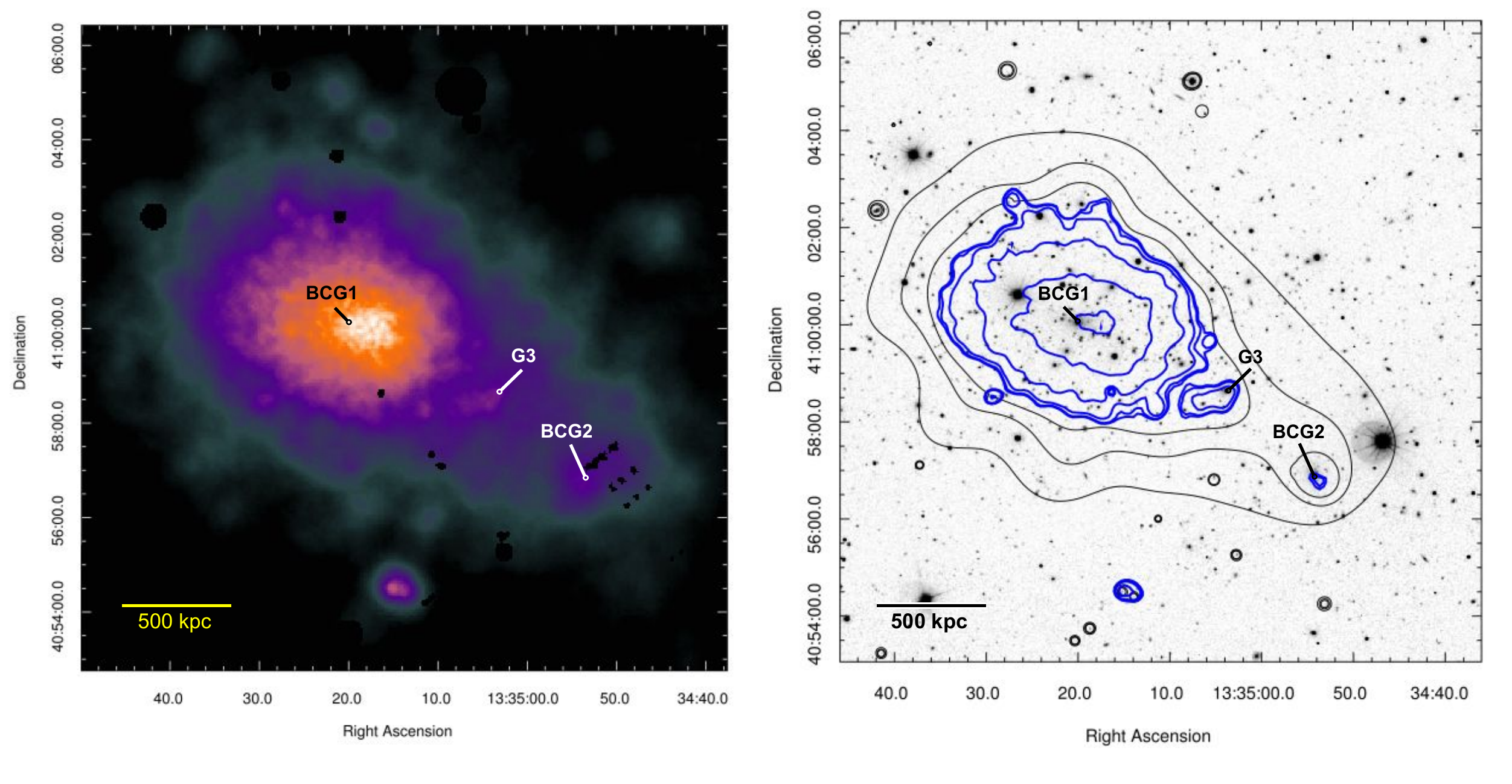}
\caption{Left panel: Adaptively smoothed 27 ks \emph{XMM-Newton} image of Abell 1763. Image is 3 Mpc (13.6$'$) on a side. Right panel: SDSS-r image of Abell 1763 covering the same field of view.  Overlaid in blue are \emph{XMM-Newton} contours with limits chosen to highlight the extension of gas trailing to the east of G3. Black contours are those of the lower surface brightness regions of the Chandra image in Figure \ref{fig:csmooth3Mpc}. }
\label{fig:XMMsdss}
\end{center}
\end{figure}

Prominent in both the observation of Abell 1763 and the simulation of \citet{Zuhone11} is the extension of emission to the east of the BCG2 subcluster. Close inspection of Figure \ref{fig:SCcon} shows that the brightest portion of the extension is nearly coincident with the third-brightest galaxy in the system (G3, r$_{sdss}$=17.27).  The location of its brightness peak is $\sim$ 50 kpc to the southeast of the galaxy.  An \emph{XMM-Newton} archival image of Abell 1763 (27 ks, Figure \ref{fig:XMMsdss}) shows a distinct region of emission extending $\sim$ 300 kpc ESE of G3, with the galaxy being located inside the feature's western edge. Based on the average ICM density at this radius and the estimated volume of the G3 extension (modeled as a cylinder of r = 20 kpc, h = 300 kpc), the mass of the gas feature is roughly $\sim 1.4 \times 10^{11}$ M$_{\odot}$.  Such a value is more typical of the core of a galaxy group \citep{Sun03,Sun04} than the ISM halo of an individual galaxy.  This suggests that G3 and the G3 extension may be the dominant galaxy and disrupted gas core of an infalling system separate from the BCG2 cluster.  

Recently, \citet{Haines18} published results from an \emph{XMM/Subaru} study probing the presence of infalling systems upon high-mass primaries at intermediate redshift (0.1 $<$ z $<$ 0.3). They identify both the BCG2 and G3 subsystems as groups descending into the Abell 1763 gravitational well. Adhering to the L$_{X}$-M$_{200}$ relation of \citet{Leauthaud10}, they determine group masses based on the luminosity calculated from a region spanning a radius equal to r$_{200}$ of the subcluster. The value of r$_{200}$ is estimated based on observed X-ray flux following the method outlined in \citet{Finoguenov07}. They arrive at group masses on the order of 10$^{14}$ M$_{\odot}$ for both the BCG2 subcluster and G3 system\footnote{The 1.2 Mpc projected separation between BCG2 subcluster and the primary cluster likely results in minimal contamination from the larger system's extended ICM when determining values for r$_{200}$ and L$_{X}$. Given this, the mass estimate for the BCG2 subcluster (2.1 $\times$ 10$^{14}$ M$_{\odot}$) is likely reliable. The mass estimate for the G3 system (1.4 $\times$ 10$^{14}$ M$_{\odot}$), however, should only serve as an upper limit, as its proximity to the primary in the plane of the sky ensures that a significant amount of ICM emission is projected onto the region occupied by the G3 system from which r$_{200}$ and L$_{X}$ is determined.}.  

The spatial distribution of the G3 trail suggests a counterclockwise infall trajectory similar to that presented for the BCG2 system. Given their similar redshifts (BCG2: z=0.238, G3: z=0.237, $\Delta$v=245 km s$^{-1}$)  the two subsystems may have descended into the primary system together as an extended or bimodal subcluster.  The dynamical analysis outlined in Section \ref{sec:velocities} suggests the BCG2 system is the primary candidate responsible for initiating spiral formation. However, spiral formation is not usually accompanied by SCC destruction. It may be the subsequent close passage of the G3 system that fully displaced BCG1 from the western peak, accelerating core disruption. The first off-axis merger with the BCG2 system would have initiated sloshing, slightly displacing the SCC from BCG1 and the DM peak. Once displaced from the cluster potential well, the SCC would have been more vulnerable to disruption by the off-axis merger with the G3 system. Combined effects of multiple minor mergers producing conditions capable of SCC destruction has been observed in the simulations of \citet{RS01} and \citet{NB99} and may be contributing to the high peculiar velocity of BCG1.

\subsubsection{The G3 System as Primary Perturber}
Alternatively, the possibility exists that the G3 system is the sole perturber.  In such a scenario, the G3 system would be the primary driver of spiral formation, SCC disruption, and high peculiar velocity of BCG1. If this were the case, the trajectory of the BCG2 system may not have included a previous pass south of the cluster. The system could be falling in from the southwest, currently in a pre-core crossing epoch. In either case, Abell 1763 is one of a small number of spiral-hosting clusters where the perturbing system is identified in X-ray and optical observations \citep[see also Abell 1644,][]{Johnson10}. To date, it is the only such system where spiral formation is observed in conjunction with SCC destruction.

\section{CONCLUSION}\label{sec:conc}

The discovery and understanding of gas-sloshing spirals ranks among the great achievements in our current era of X-ray astronomy.  Their presence reveals that many seemingly relaxed SCC clusters undergo gentle off-axis mergers that set the core in motion about the DM peak, without significantly disturbing the SCC itself.  This paper examined Abell 1763, which in contrast to the gas-sloshing paradigm, exhibits a cluster-wide spiral, but no SCC. As a spiral is understood to form from a sloshing SCC, we can assume that the WCC at the core of Abell 1763 had been in an SCC state at the initiation of sloshing. Its dynamical evolution driven by the ongoing off-axis merger has led to its current remnant CC status.

Abell 1763 appears to be the site of ongoing infall by multiple subclusters. To the southwest of the primary cluster lies a gaseous secondary system, dominated by BCG2. Dynamical analysis shows that the two systems are moving with a relative velocity of 1812 km s$^{-1}$. Given the location of the BCG2 subcluster and its orientation with respect to the spiral, it is identified as a likely candidate for initiating spiral formation. Within this scenario, the BCG2 system is on a counterclockwise path, having recently passed south of the primary cluster. Serving as indicators of the ICM velocity fields in their vicinities, WATs are hosted by both BCG1 and BCG2. Their bent lobes are oriented in a direction consistent with the described merger configuration.

The third-ranked galaxy (G3) is associated with an elongated X-ray feature and identified as the dominant galaxy in an infalling system of comparable mass to the BCG2 subcluster \citep{Haines18}. To what degree the G3 system is contributing to (or responsible for) spiral formation and SCC destruction is unclear. The high peculiar velocity that is observed for BCG1 \textbf{(v$_{pec}$ $\sim$  - 650 km s$^{-1}$)} may be attributed to this close pass of the G3 system, but it may instead be a complex dynamical consequence of the significant merger occurring with the BCG2 system.

The SCC destruction scenario depicted in this paper is characterized by off-axis infall of high-velocity systems, i.e. a dynamical environment defined by high angular momentum. This is unusual as off-axis mergers are not typically invoked as a mechanism responsible for the SCC to NCC transition.  Within the cluster merger model, SCC destruction is thought to occur almost exclusively in head-on, major mergers \citep[e.g.][]{Hahn17}. In disagreement with this, Abell 1763 and the simulations of \citet{Zuhone11}, show that off-axis mergers are indeed capable of destroying SCCs. The Abell 1763 system represents a snapshot of an off-axis merger capable of producing a remnant CC. However, the \citet{Zuhone11} simulations show that as mergers of this nature progress, the core fully transforms to an NCC state. The angular momentum present in the system ensures the core's ultimate fate, while the presence of the large-scale spiral reveals its pre-merger SCC identity. The cluster is by all accounts a transitional system, which when examined, provides valuable insight into the dynamically driven transformation of an SCC into an NCC. Knowing this, the search for additional WCC clusters hosting gas-sloshing spirals should serve as a useful avenue for exploring galaxy cluster evolution and clarifying the role that off-axis cluster mergers play in populating the NCC class.

\emph{Acknowledgments:} Basic research in radio astronomy at the US Naval Research Laboratory is supported by 6.1 Base funding.  E.M.D. would like to thank R. Cardero and M. Douglass for helpful discussions while preparing the manuscript.

\bibliography{references}

\begin{table}
\centering
\begin{tabular}{lllllllll}
Fit       &  A$_{1}$  &  $\mu_1$  &  $\sigma_1$   &  A$_{2}$  &  $\mu_2$  &  $\sigma_2$   & reduced $\chi^2$ & d.o.f. \\
\hline \hline
\textbf{one-component}    & 19.1 $_{- 2.89}^{+2.78}$ & 69656 $_{- 237}^{+253}$   & 1562 $_{- 252}^{+353}$&-&-&-&0.68&8\\
\textbf{two-component}  & 19.4 $_{- 3.95}^{+5.10}$ & 69257$_{- 239}^{+276}$   & 1151 $_{- 436}^{+431}$&12.2 $_{- 5.73}^{+5.73}$&71069 $_{-288}^{+255}$&500 &0.91&6  \\

\end{tabular}
\caption{Fitted Values for Single- and Double-Gaussian Fits to 137 cluster Members.}\label{tab:vels}

\end{table}

\end{document}